\RequirePackage[l2tabu, orthodox]{nag}
\documentclass[
    final, 
	fontsize=10.5pt,	
	paper=a4, 
	DIV=10,
	paper=portrait, 
    pagesize=auto,              
    version=last]{scrartcl}
    
\usepackage[
   centertags, 
   sumlimits,  
   intlimits,  
   namelimits, 
   fleqn,      
]{amsmath}

\usepackage[T1]{fontenc}
\usepackage[utf8]{inputenc}
\usepackage[english]{babel}
\usepackage[
  dvipsnames, 
  table,      
  hyperref,   
  fixinclude, 
]{xcolor}

\usepackage{longtable}
\usepackage{graphicx}
\usepackage{siunitx} 
\usepackage[version=3, arrows=pgf-filled]{mhchem} 
\usepackage{booktabs}
\usepackage{multirow, bigstrut}
\usepackage{multicol}
\usepackage{amssymb} 
\usepackage[figurewithin=none, tablewithin=none, margin=10pt,font=small,textfont=it,labelfont=bf]{caption}
\usepackage[section]{placeins}
\usepackage[
    url=false,
    style=numeric-comp,
    sorting=none,
    firstinits=true,
    maxcitenames=3,
    minbibnames=9,
    maxbibnames=10,
    eprint=false,
    bibencoding=utf8,
]{biblatex}

\usepackage[
    hyperindex=true,    
    hyperfootnotes=false, 
    colorlinks=true,
    linkcolor=blue,
    citecolor=OliveGreen,
    urlcolor=NavyBlue,
    bookmarks=true,
    pdfpagelabels=true
]{hyperref}

\usepackage[printonlyused]{acronym}

\hypersetup{
    pdfauthor={Marcus Kettner},
    pdftitle={Sneaking Up On The Criegee Intermediate From Below:},
    pdfsubject={Predicted Photoelectron Spectrum Of The CH_2OO^- Anion And W3-F12 Electron Affinity Of CH_2OO},
    pdfkeywords={Quantum Chemistry, Laser Chemistry, Physical Chemistry, Chemical Physics, Spectroscopy}
}

\newacro{BOA}[BOA]{Born-Oppenheimer Approximation}
\newacro{BOPES}[BO-PES]{Born-Oppenheimer Potential Energy Surface}
\newacro{HF}[HF]{Hartree-Fock}
\newacro{RHF}[RHF]{Restricted Hartree-Fock}
\newacro{ROHF}[ROHF]{Restricted Open-shell Hartree-Fock }
\newacro{UHF}[UHF]{Unrestricted Hartree-Fock}
\newacro{SCF}[SCF]{Self Consistent Field}
\newacro{MPPT}[MP-PT]{Møller-Plesset Perturbation Theory}
\newacro{RSPT}[RS-PT]{Rayleigh-Schrödinger Perturbation Theory}
\newacro{MP2}[MP2]{Second Order Møller-Plesset}
\newacro{CI}[CI]{Configuration Interaction}
\newacro{CC}[CC]{Coupled-Cluster}
\newacro{CCSD}[CCSD]{Coupled-Cluster approach including single and double exitations}
\newacro{CCSD(T)}[CCSD(T)]{Coupled-Cluster singles and doubles method with perturbative inclusion of triples}
\newacro{TDDFT}[TDDFT]{Time-dependent density functional theory}
\newacro{EOM}[EOM]{Equation-of-motion}
\newacro{CASSCF}[CASSCF]{Complete Active Space Self Consistent Field}
\newacro{LCAO}[LCAO]{Linear Combination Of Atomic Orbitals}
\newacro{DFT}[DFT]{Density Functional Theory}
\newacro{NBO}[NBO]{natural bond orbital}
\newacro{PE}[PE]{Photoelectron}
\newacro{PES}[PES]{Photoelectron Specroscopy}
\newacro{TOF}[TOF]{Time Of Flight}

\newacro{DMSO}[DMSO]{Dimethyl sulfoxide}

\begin{document}
\author{A. Karton \and M. Kettner \and D. A. Wild\thanks{Author to whom correspondence should be addressed.}}
\title{\rmfamily\Large{\textsc{Sneaking Up On The Criegee Intermediate From Below: Predicted Photoelectron Spectrum Of The \cf{CH2OO-} Anion And W3-F12 Electron Affinity Of \cf{CH2OO}}}}
\date{}
\publishers{
    \normalfont\normalsize%
    \parbox{0.6\linewidth}{%
      \begin{center}
        \textit{School of Chemistry and Biochemistry}\\
        \textit{The University of Western Australia}\\
        \textit{M310, 35 Stirling Hwy, Crawley, Australia 6009}\\
        \href{mailto:duncan.wild@uwa.edu.au}{\texttt{duncan.wild@uwa.edu.au}}
      \end{center}
    }\\
    \vfill
    \normalfont\normalsize
    \parbox{0.8\linewidth}{%
        High level ab initio calculations were undertaken on the \ce{CH2OO} anion and 
 neutral species to predict the electron affinity and anion photoelectron
 spectrum.
The electron affinity of \ce{CH2OO}, \SI{0.567}{eV}, and barrier height for 
 dissociation of \ce{CH2OO-} to \ce{O-} and \ce{CH2O}, \SI{16.5}{\kilo\joule\per\mole},
 are obtained by means of the W3-F12 thermochemical protocol.
Two major geometric differences between the anion and neutral, being the
 dihedral angle of the terminal hydrogen atoms with respect to  \ce{C-O-O}
 plane, and the  \ce{O-O} bond length, are reflected in the predicted spectrum
 as pronounced vibrational progressions.
   
    }
}
\maketitle

\section{Introduction}

Elucidation of the properties and reaction pathways of Criegee intermediates has
 gained fresh momentum recently.
These intermediates, first proposed by Criegee in 1949,
 \supercite{criegee.r.49.ozonisierung.article} are of importance in the ozonolysis   
 and breakdown of unsaturated hydrocarbons, which has been shown to be a major 
 pathway in the troposphere leading to \ce{OH} radicals and particulate 
 matter.\supercite{johnson.d.07.gas-phase.article,
 hasson.a.05.theoretical.article, gutbrod.r.97.kinetic.article,
 gutbrod.r.96.formation.article}                                   
The simplest Criegee intermediate is \ce{CH2OO} and has in previous work been 
 named formaldehyde oxide, peroxymethylene, or more generally as a carbonyl
 oxide.

The reason for the renewed activity can be attributed to a breakthrough by 
 Taatjes and co-workers who developed novel gas-phase synthetic techniques 
 for the Criegee intermediates.
They first used the chlorine atom initiated reaction with DMSO, followed by
 photolysis to form \ce{CH2OO},\supercite{taatjes.c.08.direct.article} and shortly 
 after a second route was developed whereby they photolysed  \ce{CH2I2} in the
 presence of \ce{O2}.\supercite{welz.o.12.direct.article}                            
In this second study, the authors determined rates of reaction between the
 Criegee intermediate and atmospherically important species \ce{NO}, \ce{NO2},
 \ce{SO2}, and \ce{H2O}.
The group has recently extended their initial study to the next largest Criegee
 intermediate, namely \ce{CH3CHOO}.\supercite{taatjes.c.13.direct.article}         

Since the breakthrough of this new synthetic approach there has been a burst of
 activity to characterise the Criegee intermediate by gas-phase IR spectroscopy
 in the \SIrange{800}{1500}{cm^{-1}} region and determination of the UV 
 absorption cross section and 
 photochemistry.\supercite{su.y.13.infrared.article,beames.j.12.ultraviolet.article} 
The infrared spectrum was interpreted with the aid of ab initio calculations 
 where three other possible structures for \ce{CH2O2} were computed, dioxirane, 
 methylenebis(oxy) (also known as dioxymethane), and formic acid, and compared
 with the experimental spectrum.
 
Clearly the best match between experiment and theory was with the Criegee
 intermediate, thereby providing definitive proof of its synthesis.
Lester and co-workers reported the UV absorption of the \ce{CH2OO} biradical in
 the \SIrange{320}{350}{nm} region, assigned to the \ce{B} $\leftarrow$ \ce{X}
 transition,\supercite{beames.j.12.ultraviolet.article} where the excited state      
 (\ce{B}) is purely repulsive along the \ce{O-O} coordinate.
Results from this work are extremely useful, as they identified the signature of
 the Criegee intermediate that can be applied in future laboratory or field 
 studies. 

The Criegee intermediates have received quite a deal of theoretical attention 
 over the years, with the state of play well represented in References
 \cite{fang.d.02.casscf.article}, \cite{nguyen.m.07.heats.article}       
 and \cite{vereecken.l.12.theoretical.article} and references cited therein.
Fang~et~al.\ studied the potential energy surface of the reaction of \ce{CH2} and
 \ce{O2} on the singlet and triplet surfaces using CASSCF-type        S          
 calculations.\supercite{fang.d.02.casscf.article} 
 Nguyen~et~al.\supercite{nguyen.m.07.heats.article} obtained the heat of formation  
 and ionisation energy of the Criegee intermediate at the CCSD(T)/CBS level 
 using W1 theory.\supercite{martin.j.99.towards.article}                            
It is worthwhile noting that quite recently, Dyke and co-workers used TDDFT,
 EOM-CCSD, and CASSCF methods to investigate the first few excited electronic 
 states of the simplest \ce{CH2OO} 
 intermediate.\supercite{lee.e.12.spectroscopy.article}                             
They also predicted the form of the photoelectron spectrum, i.e. {cation 
 $\leftarrow$ neutral} transitions.

In this contribution, evidence for the stability of the \ce{CH2OO-} anion from 
 high level W3-F12 calculations is provided. 
The anion is stable with respect to both autodetachment to the \ce{CH2OO}
 neutral and unimolecular dissociation to \ce{CH2O} (formaldehyde) and \ce{O-} 
 products.
We believe that it will be possible to characterise the neutral \ce{CH2OO}, and
 possibly larger, Criegee intermediates through anion photoelectron
 spectroscopy; a valuable method for mapping out the vibrational and electronic 
 states of neutral species.
These experiments are appealing as mass spectrometry can be used to isolate the
 target complex out of an ion population prior to spectroscopic interrogation.
There has been a large volume of work produced in this area from the groups of
 Lineberger, Neumark, Bowen, Wang, and Kaya, and Johnson with some 
 representative examples provided in references 
 \cite{elliott.b.08.photoelectron.article, calvi.r.07.negative.article,
 kammrath.a.06.photoelectron.article, jellinek.j.06.clusters.article,
 li.x.05.probing.article, ohara.m.02.geometric.article}.                       
Our laboratory has also recently made contributions in this 
 area.\supercite{lapere.k.11.anion.article, lapere.k.12.bromide.article, 
 lapere.k.12.anion.article}                                                     

To the best of the authors' knowledge, there have been no previous computational
 or experimental investigations of the \ce{CH2OO-} anion.
A computational study by Roos and co-workers deals with the dioxyrane, 
 dioxymethane, and the dioxymethane anion,\supercite{cantos.m.94.theoretical.article} 
 however the dioxymethane species have both oxygen atoms bound to carbon and not
 to each other, while for dioxyrane there is an \ce{O-O} bond. 
There has been experimental work undertaken on very similar species, namely the 
 alkyl peroxides \ce{CH3OO-} and \ce{CH3CH2OO-}. 
Blanksby~et~al.\ produced negative ion photoelectron spectra and determined the 
 adiabatic electron affinities of \ce{CH3OO} and \ce{CH3CH2OO} to be
 \SI{1.161+-0.005}{eV} and \SI{1.154+-0.004}{eV}
 respectively.\supercite{blanksby.s.01.negative-ion.article}                         
The spectrum of the \ce{CH3OO-} anion showed resolved vibrational progressions
 in the \ce{O-O} stretching and \ce{H3C-O-O} bending modes.
The spectrum of the \ce{CH3CH2OO-} also showed well resolved vibrational 
 progressions. 
Assignment of one progression to the \ce{O-O} stretch was clear, while a second 
 progression was assigned to a bending mode; however, it was unclear whether it
 was the \ce{C-C-O} or \ce{C-O-O} mode as these modes are predicted to lie 
 very close in frequency. 
With regard to the alkyl peroxy systems it would be remiss not to make reference
 to the work undertaken by Xu~et~al.\supercite{xu.w.08.structures.article}           
They performed an extensive computational study on the \ce{R-OO} and \ce{R-OO-}
 anion species with \ce{R} = \ce{CH3}, \ce{C2H5}, $n$-\ce{C3H7}, $n$-\ce{C4H9},
 $n$-\ce{C5H11}, $i$-\ce{C3H7}, and $t$-\ce{C4H9}.
They predicted the structures, vibrational frequencies, and electron affinities
 of the neutral species using seven different density functional or hybrid
 density functional methods.
 
In the present Letter we obtain the heat of formation and electron affinity of
 the Criegee intermediate using the recently developed W3-F12 
 theory.\supercite{karton.a.12.explicitly.article}                                   
W3-F12 represents a layered extrapolation to the relativistic, all-electron 
 CCSDT(Q)/CBS energy (complete basis set limit coupled cluster with singles, 
 doubles, triples, and quasiperturbative quadruple excitations) and shows 
 excellent performance for systems containing first-row elements (and H).
Specifically, over the \num{97} first-row systems in the W4-11 
 dataset,\supercite{karton.a.12.w4-12.article} W3-F12 attains a root mean square
 deviation (RMSD) of only \SI{0.88}{\kilo\joule\per\mole} against all-electron, relativistic
 reference atomisation energies obtained close to the full configuration
 interaction (FCI) infinite basis set limit.
In addition to the heats of formation and electron affinity of the \ce{CH2OO}
 species, we obtain the barrier height and energy for the 
 \ce{CH2OO- -> CH2O + O-} reaction using W3-F12 theory.
\section{Computational Methods}

The geometries and harmonic frequencies of the anion and neutral \ce{CH2OO}
 species were obtained at the CCSD(T)/A$'$VQZ level of theory (where A$'$V$n$Z
 indicates the combination of  Dunning's aug-cc-pV$n$Z basis set on carbon and
 oxygen and the standard cc-pV$n$Z basis set on 
 hydrogen).\supercite{dunning.t.89.gaussian.article, 
 kendall.r.96.electron.article}                                                 
The geometry and harmonic frequencies for the transition structure of the
 \ce{CH2OO- -> CH2O + O-} reaction are obtained at the CCSD(T)/A$'$VTZ level of
 theory.
The optimized geometries for all the species considered in the present work are
 given as Cartesian coordinate form in \autoref{App: optimised geometries}                         
 of the supporting information.
All the high-level ab initio calculations were performed using the 
 \textsc{Molpro} program suite,\supercite{MOLPRO} while all the density 
 functional theory (DFT) calculations were carried out with the 
 \textsc{Gaussian} 09 program suite.\supercite{g09}                                                         

The total atomisation energies at the bottom of the well (TAE$_e$)
 of the \ce{CH2OO} and \ce{CH2OO-} species are obtained by means of the W3-F12
 procedure.\supercite{karton.a.12.explicitly.article}                           
W3-F12 theory combines F12 methods \supercite{peterson.k.12.initio.article,
ten-no.s.12.explicitly.article}                                                 
 with extrapolation techniques in order to reproduce the CCSDT(Q) basis set
  limit energy.
The CCSD(T)/CBS energy is obtained from the W2-F12 theory and the post-CCSD(T)
 contributions are obtained from W3.2 theory.\supercite{karton.a.06.w4.article} 
In brief, the Hartree{\textendash}Fock component is calculated with the VQZ-F12 
 basis set (V$n$Z-F12 denotes the cc-pV$n$Z-F12 basis sets of Peterson et al.\
 which were developed for explicitly correlated
 calculations).\supercite{peterson.k.08.systematically.article}                 
Note that the complementary auxiliary basis (CABS) singles correction is
 included in the SCF energy.\supercite{knizia.g.08.explicitly.article,
adler.t.07.simple.article,noga.j.07.second.article}                             
The valence CCSD-F12 correlation energy is extrapolated from the VTZ-F12 and 
 VQZ-F12 basis sets, using the $E(L) = E_{\infty} + A/L^{\alpha}$ two-point
 extrapolation formula, with $\alpha$ = \num{3.67}.
In all of the explicitly-correlated coupled cluster calculations the diagonal,
 fixed-amplitude 3C(FIX) ansatz\supercite{knizia.g.08.explicitly.article,
ten-no.s.04.initiation.article, werner.h.07.explicitly.article}                 
 and the CCSD-F12b approximation\supercite{adler.t.07.simple.article,
knizia.g.09.simplified.article} are employed.                                   
The quasiperturbative triples, (T), corrections are obtained from standard
 CCSD(T) calculations (i.e., without inclusion of F12 terms) and scaled by the 
 factor $f = 0.987 \cdot E^{\mathrm{MP2-F12}}/E^{\mathrm{MP2}}.$
This approach has been shown to accelerate the basis set 
 convergence.\supercite{karton.a.12.explicitly.article,
 knizia.g.09.simplified.article}                                                
The higher-order connected triples, T$_3$-(T), valence correlation contribution
 is extrapolated from the cc-pVDZ and cc-pVTZ basis sets using the above
 two-point extrapolation formula with $\alpha$= 3, and the parenthetical 
 connected quadruples contribution (CCSDT(Q)--CCSDT) is calculated with the 
 cc-pVDZ basis set.\supercite{karton.a.06.w4.article}                           
The CCSD inner-shell contribution is calculated with the core-valence weighted
 correlation-consistent A$'$PWCVTZ basis set of Peterson and 
 Dunning,\supercite{peterson.k.02.accurate.article} whilst the (T) inner-shell  
 contribution is calculated with the PWCVTZ(no $f$) basis set  (where A$'$PWCVTZ
 indicates the combination of the cc-pVTZ basis set on hydrogen and the 
 aug-cc-pwCVTZ basis set on carbon, and PWCVTZ(no $f$) indicates the cc-pwCVTZ
 basis set without the $f$ functions).\supercite{karton.a.12.explicitly.article}
The scalar relativistic contribution (in the second-order 
 Douglas{\textendash}Kroll{\textendash}Hess 
 approximation\supercite{douglas.m.74.quantum.article, 
 hess.b.86.relativistic.article}) is obtained as the difference between         
 non-relativistic CCSD(T)/A$'$VDZ and relativistic CCSD(T)/A$'$VDZ-DK
 calculations (where A$'$VDZ-DK indicates the combination of the cc-pVDZ-DK
 basis set on H and aug-cc-pVDZ-DK basis set on C and 
 O).\supercite{jong.w.01.parallel.article}                                     
The atomic spin-orbit coupling terms are taken from the experimental fine 
 structure, and the diagonal Born{\textendash}Oppenheimer corrections (DBOC) are
 calculated at the HF/A$'$VTZ level of theory.
The zero-point vibrational energies (ZPVEs) are derived from the harmonic 
 frequencies (calculated at the CCSD(T)/A$'$VQZ level of theory for the
 \ce{CH2OO} and \ce{CH2OO-} species, and the CCSD(T)/A$'$VTZ level of theory 
 for the \ce{CH2O}$\cdots$\ce{O-} transition structure).

The total atomisation energies at \SI{0}{\kelvin} (TAE$_0$) are converted to a 
 heats of formation at \SI{298}{\kelvin} using the Active Thermochemical Tables 
 (ATcT)\supercite{ruscic.b.04.introduction.article, stevens.w.10.heats.article,
  ruscic.b.06.active.article}                                                   
 atomic heats of formation at \SI{0}{\kelvin} 
 (\ce{H} \SI{216.034 +- 0.001}{\kilo\joule\per\mole},
  \ce{C} \SI{711.38 +- 0.06}{\kilo\joule\per\mole},
 and \ce{O} \SI{246.844 +- 0.002}{\kilo\joule\per\mole}), and the 
 CODATA\supercite{CODATA} enthalpy functions,
 $H_{298}-H_0$, for the elemental reference states (\ce{H2(g)} =  
 \SI{8.468 +- 0.001}{\kilo\joule\per\mole} and \ce{C}($cr$,graphite) = 
 \SI{1.050 +- 0.020}{\kilo\joule\per\mole}), while the enthalpy functions for 
 the \ce{CH2OO} and \ce{CH2OO-} species are obtained within the ridged rotor 
 harmonic oscillator (RRHO) approximation from B3LYP/A$'$VTZ geometries and 
 harmonic frequencies.\supercite{becke.a.93.density-functional.article, 
 stephens.p.94.initio.article, lee.c.88.development.article}                   
Anion photoelectron spectra were simulated by determining the 
 Franck{\textendash}Condon Factors (FCFs) linking the anion and neutral 
 \ce{CH2OO} species vibrational states.
FCFs were calculated using the ezSpectrum 3.0 program which is made freely
 available by Mozhayskiy and Krylov.\supercite{ezSpectrum}                      
The program produces FCFs in either the parallel mode approximation as products
 of one-dimensional harmonic wavefunctions, or by undertaking 
 Duschinsky rotations of the normal modes between states.
Input to the code consists of the output from the ab initio calculations, being
 geometries, vibrational frequencies, and vibrational normal mode vectors.
The predicted stick spectra were convoluted with a Gaussian response function of
 width \SI{0.002}{\electronvolt} to simulate an experimental spectrum.
\section{Results \& Discussion}

\subsection{Geometries and vibrational frequencies}
\begin{figure}[h!]
    \caption{The \cf{CH2OO} anion and neutral species. Bond lengths are from 
             CCSD(T)/A$'$VQZ calculations. Full geometric data are presented in
             \autoref{Dataset}.
            }
    \centering
    \includegraphics[width=0.8\linewidth]{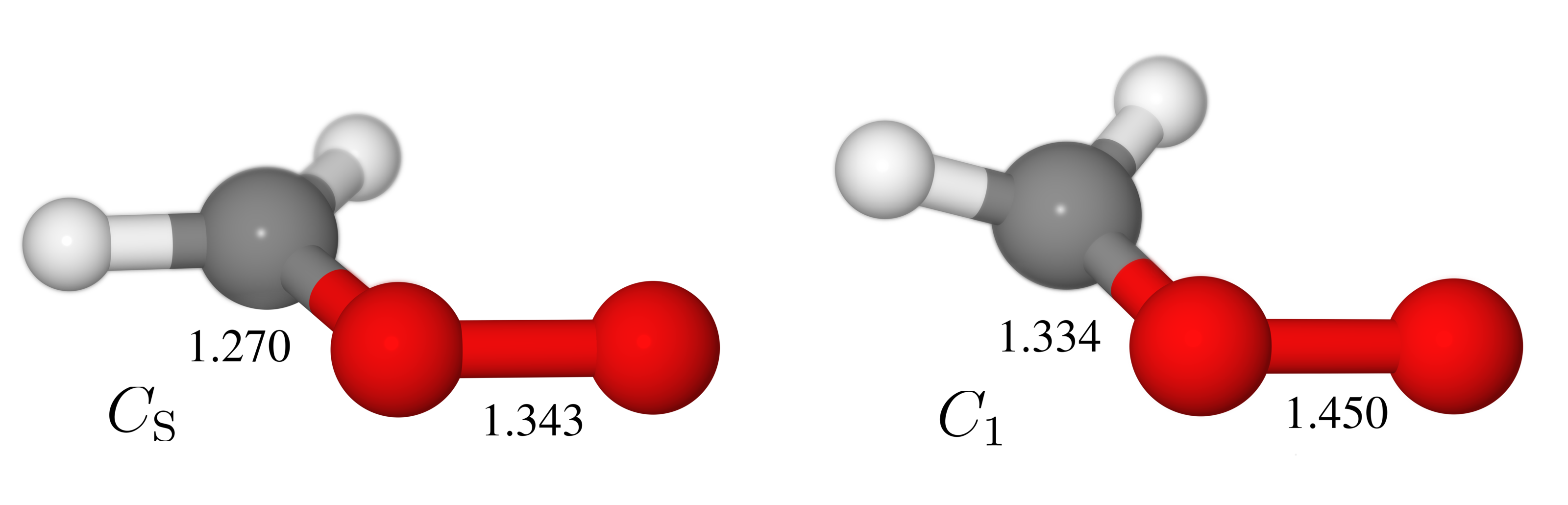}
    \label{Comparisson}
\end{figure}

The geometry of the neutral Criegee intermediate is well known from previous 
 high level calculations (see Reference \cite{nguyen.m.07.heats.article} and    
 References cited therein).
It has been shown that the \ce{C-O} and \ce{O-O} bond lengths are sensitive to 
 the level of theory used is, and it is noted that the values predicted from 
 CCSD(T)/A$'$VQZ calculations are in line with those reported in references 
 \cite{fang.d.02.casscf.article} and \cite{nguyen.m.07.heats.article}.          
Our full data set is provided in \autoref{Dataset}, and a visual comparison of the 
 two species is provided in \autoref{Comparisson}.
We predict values of \SI{1.270}{\angstrom} and \SI{1.343}{\angstrom} for the 
 \ce{C-O} and \ce{O-O} bond lengths respectively, which is in very good
 agreement with the CCSD(T)/AVTZ calculations of Nguyen et al.\  
 (\SI{1.275}{\angstrom} and \SI{1.375}{\angstrom} from reference 
 \cite{nguyen.m.07.heats.article}).
 
The agreement is also very good between our results and those from Fang et al.
 \supercite{fang.d.02.casscf.article} CAS-(8,6)+1+2/cc-pVDZ calculations
 which result in \ce{C-O} and \ce{O-O} bond lengths of \SI{1.280}{\angstrom} and 
 \SI{1.322}{\angstrom} respectively.
Our CCSD(T)/A$'$VQZ harmonic vibration frequencies are also in good agreement
 with those reported previously as shown in \autoref{Dataset}.
\begin{table}[htb!]
    \caption{Computed geometric parameters of the \ce{CH2OO} anion and neutral 
    species at the CCSD(T) level of theory, with A$'$VQZ or A$'$VTZ basis sets.}
    \label{Dataset}
    \centering\small
    \begin{tabular}{llSSS}
        \toprule
                                &                   &{\textbf{Anion}}               &  {\textbf{Anion TS}}         & {\textbf{Neutral}}\\
            Basis Set           &                   & {CCSD(T)/A$'$VQZ}             &  {CCSD(T)/A$'$VTZ}           &  {CCSD(T)/A$'$VQZ}\\\midrule

            $r$(\ce{C-H})       & [\si{\angstrom}]  &  {1.086, 1.090$^{\dagger}$}   & {1.093, 1.097$^{\dagger}$}   &  {1.080, 1.082$^{\dagger}$}\\
            $r$(\ce{C-O})       & [\si{\angstrom}]  &  1.334    &  1.317  &  1.270\\
            $r$(\ce{O-O})       & [\si{\angstrom}]  &  1.450    &  1.589  &  1.343\\
            $\theta$(\ce{H-C-H})& [\si{\degree}]       &  121.1    &  118.2  &  126.5\\
            $\theta$(\ce{H-C-O})& [\si{\degree}]       &  {116.4, 113.0$^{\dagger}$}   & {118.2, 119.2$^{\dagger}$}   &  {118.7, 114.9$^{\dagger}$}\\
            $\theta$(\ce{C-O-O})& [\si{\degree}]       &  111.7    &  100.7  &  117.9\\
            $\phi$(\ce{H-C-H-O})& [\si{\degree}]       &  {-145.2, 144.1$^{\dagger}$}  & {-155.9, -156.1$^{\dagger}$} &  -180.0\\
            $\phi$(\ce{H-C-O-O})& [\si{\degree}]       &  {-17.9, -164.8$^{\dagger}$}  & {57.1, -98.8$^{\dagger}$}    &  {0.0, 180.0$^{\dagger}$}\\\bottomrule
            \multicolumn{5}{l}{\parbox{3mm}{$\dagger$\\ }\parbox{0.8\linewidth}{Where two values are present, the first corresponds to the hydrogen closest to the terminal oxygen.}}
    \end{tabular}
\end{table}

To the best of our knowledge, the geometry of the analogous Criegee anion
 \ce{CH2OO-} and the geometrical parameters (bond lengths $r$, bond angles
 $\theta$, and torsional angles $\phi$) at the CCSD(T)/A$'$VQZ level of theory 
 are presented here for the first time.
In essence, the structure of the anion is similar to the neutral, as can be seen
 in \autoref{Comparisson}, however features longer \ce{C-O} and \ce{O-O} bond 
 lengths. 
The \ce{C-O} bond length increases by \SI{0.064}{\angstrom}, while the O-O bond
 length increases by \SI{0.107}{\angstrom}.
In addition to these two structural differences, the anion species is 
 no longer of $C_{\mathrm{s}}$ symmetry as the hydrogen atoms have moved out of 
 the plane defined by the \ce{C-O-O} atoms.
As presented in \autoref{Dataset}, the \ce{H-C-O-O} dihedral angles are \ang{-17.9}
 and \ang{-164.8} for the Criegee anion, whereas in the neutral species the 
 values are \ang{0} and \ang{180}.

In order to rationalise the differences in anion and neutral geometries we have
 performed \ac{NBO} calculations to determine the population of the bonding,
 anti-bonding, and lone pair natural orbitals.\supercite{NBO6}
These data, in terms of orbital occupancies, are provided in the supplementary
 information (\autoref{App: NBO}).
For the anion species increased electron density is predicted in the lone pair
 orbitals of the carbon, and for the oxygen bound to carbon.
The pyramidal nature of the anion complex, compared with the planar neutral,
 can be rationalised as the population in the carbon lone pair orbital.
The \ac{NBO} calculations also reveal decreased electron density in \ce{C-O}
 and \ce{O-O} bonding orbitals for the anion, leading to the increase in the
 bond lengths. 
There is also a marked decrease in the population of the \ce{C-O} anti-bonding
 orbital, whereas the \ce{O-O} anti-bonding orbital occupancy barely changes.
This explains the larger change in the \ce{O-O} bond length compared with the
 \ce{C-O} bond length.

The vibrational modes of the neutral species have been ordered according to the 
 usual numbering scheme appropriate for a $C_s$ symmetry molecule (wavenumber 
 descending in $a'$, and then descending in $a''$ symmetry). 
For ease of comparison between the anion and neutral modes, we have adopted the 
 same order for the anion species even though they are all of $a$ symmetry. 
In any event, the mode description for each mode is provided in the table.
We note that the increased \ce{O-O} bond length in the anion species is
 reflected in a reduction of the frequency of the \ce{O-O} stretching mode,
  $\nu_6$, from \SIrange{945}{776}{cm^{-1}}, similarly the \ce{C-O} bond length
  increase leads to a reduction in vibrational frequency of modes $\nu_3$ and 
  $\nu_4$.
The \ce{CH2} wagging mode has also decreased in frequency in the anion species.
\begin{table}[htb!]
    \caption{Computed vibrational data of the \ce{CH2OO} anion and neutral species at the
CCSD(T)/A$'$VQZ level, and comparison with recent published work.}
    \label{Dataset2}
    \centering\small
    \begin{tabular}{llcccccl}
        \toprule
                    &              & Anion    & \multicolumn{4}{c}{Neutral}  & Mode description \\\cmidrule(r){3-3} \cmidrule(r){4-7} \cmidrule(r){8-8}
\multicolumn{2}{l}{Method}         & CCSD(T)  & CCSD(T) & CCSD(T) & \multicolumn{2}{c}{NEVPT2}& \\
\multicolumn{2}{l}{Basis Set}      & A$'$VQZ  & A$'$VQZ & AVTZ    & \multicolumn{2}{c}{AVDZ}  & \\
\multicolumn{2}{l}{Approx.}  & Har.     & Har.      & Har.      & Anh.  & Har.      & \\\cmidrule(r){1-2}\cmidrule(r){3-3} \cmidrule(r){4-4} \cmidrule(r){5-5} \cmidrule(r){6-7} \cmidrule(r){8-8}
$\nu_1$             & [\si{cm^{-1}}] & 3185 $a$ & 3302 $a'$ & 3290 $a'$ & 3370 $a'$ & 3149 $a'$ & Asymmetric \ce{CH} stretch\\
$\nu_2$             & [\si{cm^{-1}}] & 3048 $a$ & 3140 $a'$ & 3137 $a'$ & 3197 $a'$ & 3030 $a'$ & Symmetric \ce{CH} stretch\\
$\nu_3$             & [\si{cm^{-1}}] & 1422 $a$ & 1489 $a'$ & 1483 $a'$ & 1500 $a'$ & 1458 $a'$ & \ce{CH2} scissor/\ce{CO} stretch\\
$\nu_4$             & [\si{cm^{-1}}] & 1252 $a$ & 1316 $a'$ & 1306 $a'$ & 1338 $a'$ & 1302 $a'$ & \ce{CO} stretch/\ce{CH2} scissor\\
$\nu_5$             & [\si{cm^{-1}}] & 1164 $a$ & 1239 $a'$ & 1231 $a'$ & 1235 $a'$ & 1220 $a'$ & \ce{CH2} rocking\\
$\nu_6$             & [\si{cm^{-1}}] &  776 $a$ &  945 $a'$ &  935 $a'$ &  916 $a'$ &  892 $a'$ & \ce{OO} stretch\\
$\nu_7$             & [\si{cm^{-1}}] &  466 $a$ &  532 $a'$ &  529 $a'$ &  536 $a'$ &  530 $a'$ & \ce{COO} deformation\\
$^{\dagger}\nu_8$   & [\si{cm^{-1}}] &  601 $a$ &  876 $a''$&  862 $a''$&  856 $a''$&  853 $a''$& \ce{CH2} wagging\\
$^{\dagger}\nu_9$   & [\si{cm^{-1}}] &  316 $a$ &  651 $a''$&  632 $a''$&  620 $a''$&  606 $a''$& \ce{CH2} twisting\\
\multicolumn{2}{l}{Reference} & This work & This work & \cite{nguyen.m.07.heats.article} & \multicolumn{2}{c}{\cite{su.y.13.infrared.article}}   & as in \cite{su.y.13.infrared.article}\\\bottomrule
            \multicolumn{8}{l}{\parbox{3mm}{$\dagger$\\ }\parbox{0.8\linewidth}{The ordering of the modes of the anion 
            has been changed to reflect that of the neutral, for direct comparison.}}
    \end{tabular}
\end{table}

\subsection{W3-F12 heat of formation and electron affinity of the Criegee intermediate}

The components of the W3-F12 total atomisation energies for the \ce{CH2OO} and 
 \ce{CH2OO-} species are given in \autoref{W3-F12 Data}.
At the W2-F12 level, the relativistic, all-electron CCSD(T) contributions to 
 TAE$_0$ add up to \SI{1520.9}{\kilo\joule\per\mole} for \ce{CH2OO}, and 
 \SI{1581.5}{\kilo\joule\per\mole} for \ce{CH2OO-}.
The TAE$_0$ for the Criegee intermediate is higher than the W1 value of 
 Nguyen~et~al.\ by 
 \SI{1.8}{\kilo\joule\per\mole}.\supercite{nguyen.m.07.heats.article}
The lion's share of the difference comes from the valence CCSD(T) components
 which are closer to the basis set limit in W2-F12 theory 
 (\autoref{App: Comparrison W2-F12,W1a} of the Supporting Information).
The reminder of the difference comes mostly from our better geometry and ZPVE
 and from the DBOC contribution, which is not included in the W1 values.
 
We now turn our attention to the post-CCSD(T) contributions to the TAE. 
Table S3 \autoref{App: Diagnostics}
 of the Supporting Information provides a number of a priori
 diagnostics for the importance of nondynamical correlation effects, namely the 
 percentage of the total atomisation energy accounted for by the SCF and (T) 
 triples contributions from W3-F12 theory\supercite{karton.a.12.w4-12.article, 
 karton.a.06.w4.article} (as well as the coupled                                
 cluster T$_1$ and D$_1$ diagnostics).\supercite{lee.t.03.comparison.article,
 leininger.m.00.new.article, lee.t.89.theoretical.article}                      
The \ce{CH2OO} neutral and anion species considered in the present study exhibit
 mild-to-moderate nondynamical correlation effects; \SIrange{53}{55}{\percent}
 of the atomisation energy is accounted for at the SCF level, and 
 \SIrange{3.8}{4.6}{\percent} by the (T) triples.  
The T$_1$ diagnostics of above \num{0.02} (namely, \numrange{0.034}{0.045}) and 
 D$_1$ diagnostics of above \num{0.05} (\numrange{0.122}{0.176}) also indicate 
 that post-CCSD(T) excitations may have nontrivial contributions.
The generally good performance of the CCSD(T)/CBS level of theory in
 computational thermochemistry can typically be attributed to the large degree
 of cancellation between higher-order triples contributions, T$_3$-(T), and 
 post-CCSDT contributions. 
For systems dominated by dynamical correlation, these contributions are of
 similar magnitudes, however, the T$_3$-(T) excitations tend to universally
 decrease the atomisation energies whereas the post-CCSDT excitations tend to
 universally increase the atomisation energies.
In this regard, we find that for the \ce{CH2OO-} anion there is a significant
 degree of cancellation between the T$_3$-(T) contribution 
 (\SI{-2.7}{\kilo\joule\per\mole}) and the (Q) contribution 
 (\SI{+4.5}{\kilo\joule\per\mole}).
Resulting in a post-CCSD(T) contribution of \SI{1.8}{\kilo\joule\per\mole}.
However, for the Criegee intermediate there is significantly poorer cancellation
 between the T$_3$-(T) (\SI{-3.4}{\kilo\joule\per\mole}) and (Q) contributions 
 (\SI{+9.8}{\kilo\joule\per\mole}).
Therefore, the post-CCSD(T) contributions increase the atomisation energy of
 \ce{CH2OO} by as much as \SI{6.4}{\kilo\joule\per\mole}. We note that the 
 inclusion of  higher-order quadruple contributions, T$_4$-(Q), is likely to 
 reduce the magnitude of the connected quadruple excitations, and therefore our
 CCSDT(Q)/CBS values should be regarded as upper limits of the TAEs. 
\begin{table}[htb!]
    \caption{Component breakdown of the W3-F12 total atomisation energies and
             heats of formation of the \ce{CH2OO} neutral and anion species,
             electron affinity of \ce{CH2OO}, and barrier height for the 
             \ce{CH2OO- -> CH2O + O-} reaction (in \si{\kilo\joule\per\mole}).}
    \label{W3-F12 Data}
    \centering\small
    \begin{tabular}{lSSSS}
        \toprule
        & {\ce{CH2OO}} & {\ce{CH2OO-}} & {$E_{\mathrm{EA}}^{\dagger}$} & {$E_{\mathrm{BH}}^{\ast}$}\\\midrule
        SCF                         & 873.4 & 903.1  & 29.7     & 34.0  \\
        CCSD                        & 652.0 & 685.8  & 33.7     & -4.6  \\
        (T)                         & 74.4  & 62.9   & -11.5    & -5.8  \\
        T$_3$--(T)                   & -3.4 & -2.7   & 0.8      & -1.9  \\
        (Q)                         & 9.8   & 4.5    & -5.3     & -1.3  \\
        Inner-Shell                 & 5.5   & 5.2    & -0.3     & 0.5   \\
        Scalar Relativistic         & -1.9  & -2.2   & -0.3     & -0.2  \\
        Spin-Orbit                  & -2.2  & -2.2   & 0.0      & 0.0   \\
        DBOC                        & 0.3   & 0.1    & -0.1     & 0.1   \\
        TAE$_e$                     & 1608.0& 1654.6 & 46.6     & 20.7  \\
        ZPVE                        & 80.7  & 71.3   & -9.4     & 4.3   \\
        TAE$_0$                     &1527.3 & 1583.3 & 56.1     & 16.4  \\
        $\Delta H_{\mathrm{f},0}$   & 109.9 & 53.8   & 56.1     & 16.4  \\
        $\Delta H_{\mathrm{f},298}$ & 102.8 & 48.1   & 54.7     & 16.5  \\\bottomrule
        \multicolumn{5}{l}{$\dagger$ Energy for the \ce{CH2OO- -> CH2OO + e-} reaction.}\\
        \multicolumn{5}{l}{$\ast$ Barrier height for the \ce{CH2OO- -> CH2O + O-} reaction.}
    \end{tabular}
\end{table}

Overall, our best relativistic, all-electron CCSDT(Q)/CBS atomisation energies
 (\autoref{W3-F12 Data}) are \SI{1527.3}{\kilo\joule\per\mole} (\ce{CH2OO}) and 
 \SI{1583.3}{\kilo\joule\per\mole} (\ce{CH2OO-}).
These correspond to heats of formation at \SI{0}{\kelvin} of 
 \SI{109.9}{\kilo\joule\per\mole}  (\ce{CH2OO}) and 
 \SI{53.8}{\kilo\joule\per\mole} (\ce{CH2OO-}), and 
 heats of formation at  
 \SI{298}{\kelvin} of \SI{102.8}{\kilo\joule\per\mole} (\ce{CH2OO}) and 
 \SI{48.1}{\kilo\joule\per\mole} (\ce{CH2OO-}).
In accordance with the large post-CCSD(T) contributions for the Criegee
 intermediate our W3-F12 heats of formation are lower than the W1 values of
 Nguyen et al.\supercite{nguyen.m.07.heats.article} ($\Delta H_{\mathrm{f},0}=$
 \SI{113.0}{\kilo\joule\per\mole} and 
 $\Delta H_{\mathrm{f},298}=$ \SI{105.9}{\kilo\joule\per\mole}).
 
Using the W3-F12 heats of formation for the \ce{CH2OO} neutral and anion species
 electron affinities for the Criegee intermediate of 
 \SI{56.1}{\kilo\joule\per\mole} at \SI{0}{\kelvin} and 
 \SI{54.7}{\kilo\joule\per\mole} at \SI{298}{\kelvin} were obtained.
It is noted that the post-CCSD(T) contributions to the electron affinity add up 
 to as much as \SI{+4.5}{\kilo\joule\per\mole}.
 
\subsection{Stability of \cf{CH2OO-} with respect to dissociation}

\begin{figure}[htb!]
    \caption{Reaction profile illustrating the dissociation channel of the 
    Criegee anion \ce{CH2OO-} to the products \ce{CH2O + O-} calculated at the 
    W3-F12 level.}
    \centering
    \includegraphics[width=0.5\linewidth]{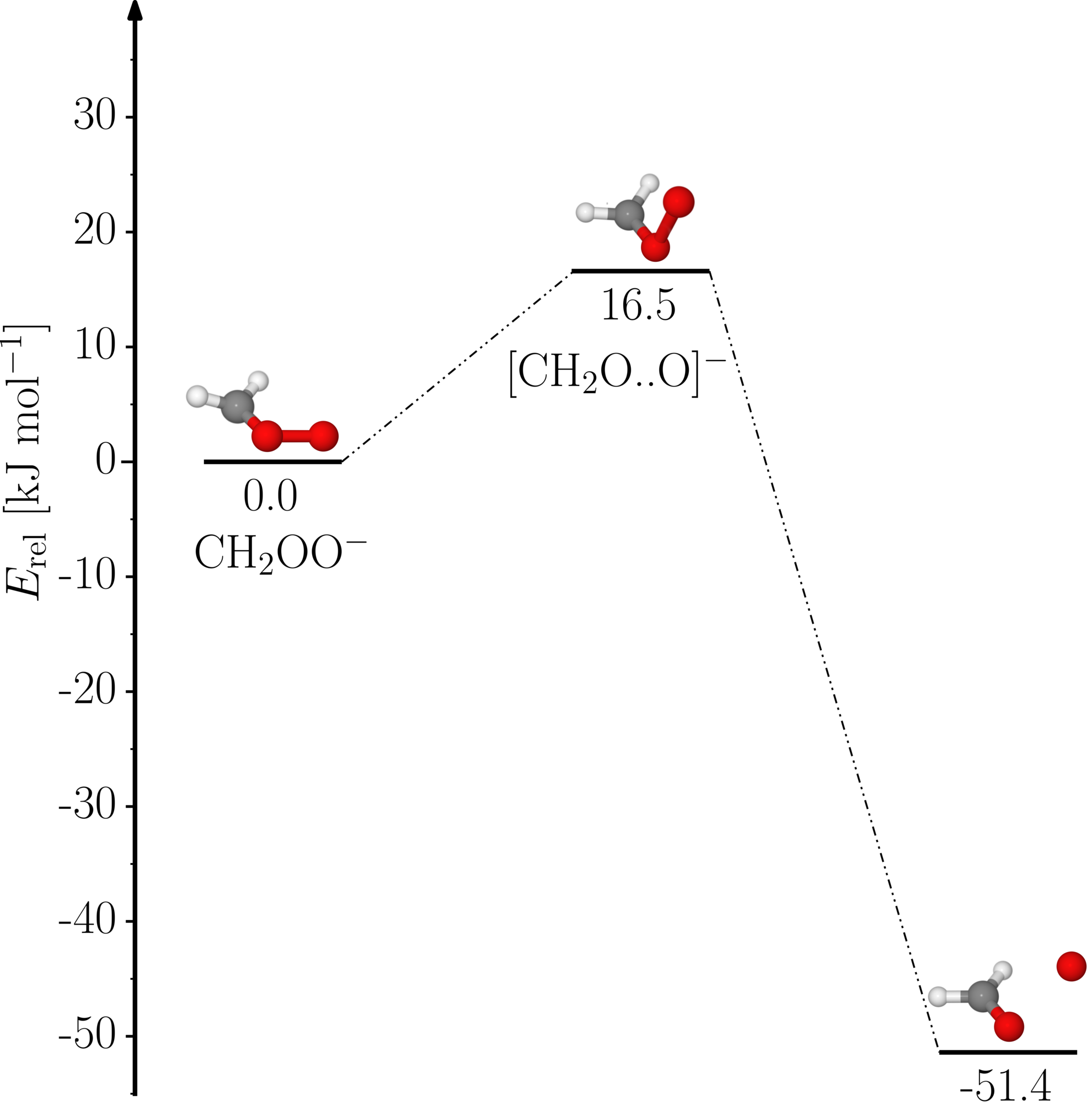}
    \label{Reaction Profile}
\end{figure} 

In order to determine the stability of the Criegee anion species we modelled the
 reaction profile for dissociation to the products formaldehyde and oxide.
This was achieved by optimising the geometry of the transition state (TS)
 linking the reactant \ce{CH2OO-} and the products \ce{CH2O} and \ce{O-}
 species.
The W3-F12 procedure was applied to the TS, anion, and products with the results
 shown in \autoref{Reaction Profile}.
It is clear that an appreciable barrier to dissociation exists of 
 \SI{16.5}{\kilo\joule\per\mole}, and therefore should a synthetic route to the
 anion be found it is likely that it should survive long enough for 
 interrogation.
 
\subsection{Predicted anion photoelectron spectrum of \cf{CH2OO-}}

\begin{figure}[htb!]
    \caption{Predicted anion photoelectron stick spectra (red marks the
            pure progressions).}
    \centering
    \includegraphics[width=1.0\linewidth]{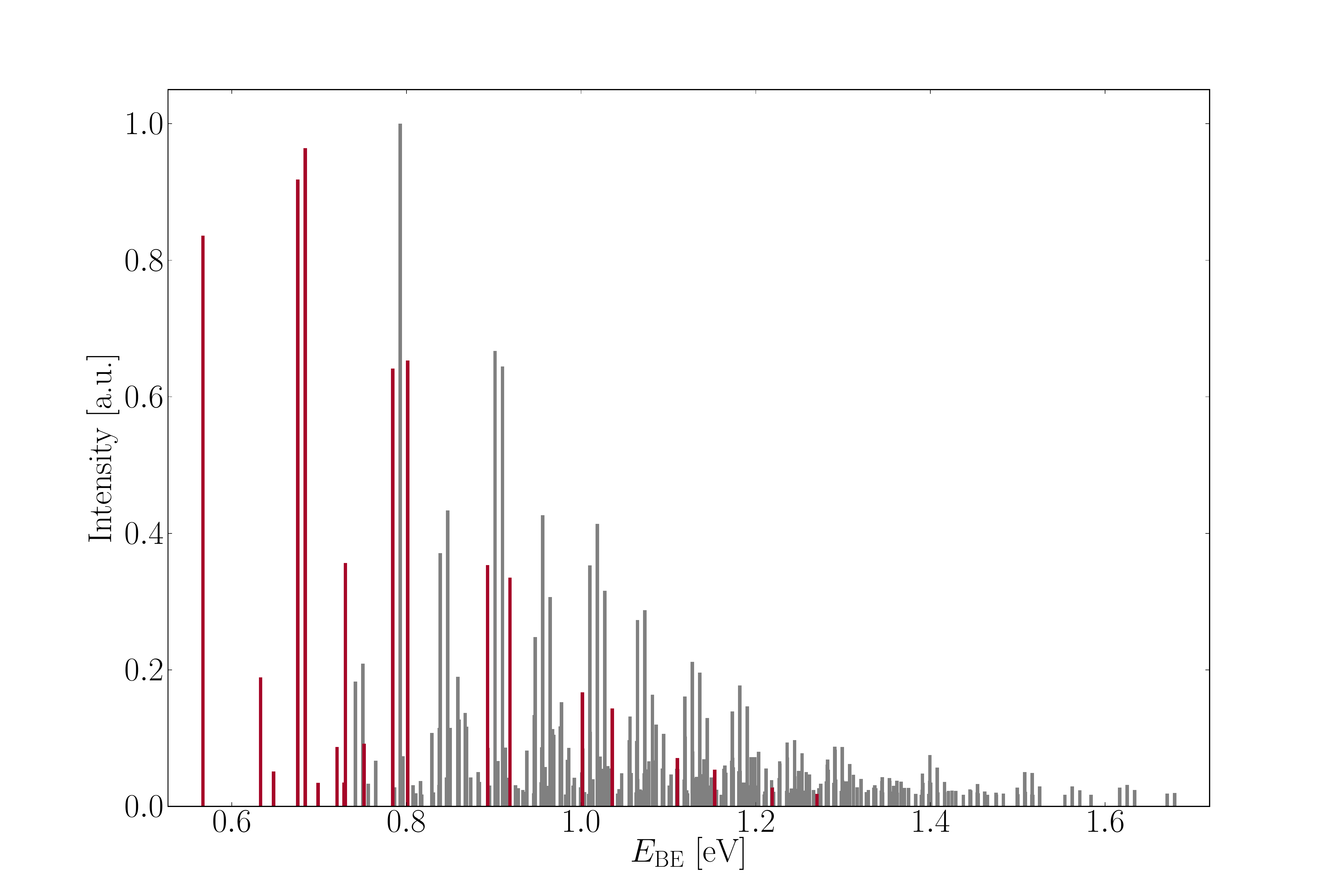}
    \label{Stick-Spectrum}
\end{figure}

Armed with the CCSD(T)/A$'$VQZ geometries, vibrational frequencies, and normal
 mode vectors, we are in a position to predict the form of the anion
 photoelectron spectrum. 
Therefore the ezSpectrum 3.0 code was employed to simulate the vibronic
 transitions at a temperature of \SI{10}{\kelvin}, which is appropriate for 
 species entrained in a molecular beam produced via supersonic expansion. 
Up to \num{10} quanta were allowed in each excited state vibrational mode 
 (i.e.\ the modes of the neutral \ce{CH2OO} species).
In \autoref{Stick-Spectrum}, a stick spectrum is presented which is the 
 result of applying the ezSpectrum code with the Duschinsky approach.
We also performed simulations using the parallel mode approximation, however,
 the differences between the forms of the two predicted spectra were 
 insignificant and do not warrant further discussion.
The grey part of the spectrum marks the combination bands, whereas the red part
 represents the pure progressions; together they form the  fully predicted
 spectrum.
Data pertaining to this spectrum, including predicted line positions, 
 intensities, Franck-Condon Factors, and assignments, are provided in the 
 supplementary information (\autoref{App: Duschinski Data} and  
 \autoref{App: Parallel Data}) accompanying this Letter.

The majority of the pure progressions (marked red in \autoref{Stick-Spectrum})
 are for the modes $\nu_6$ and $\nu_8$ which correspond to the \ce{O-O}
 stretching and \ce{CH2} wagging modes respectively.
The fact that these two modes display the longest progressions is consistent
 with the geometry changes observed between the anion and neutral species.
Referring to the data in \autoref{Dataset}, the \ce{O-O} bond length is predicted
 to decrease by \SI{0.107}{\angstrom}, while the two hydrogen atoms move into 
 the \ce{C-O-O} plane, with the pair of \ce{H-C-O-O} dihedral angles changing
 from \ang{-17.9} and \ang{-164.8} to \ang{0.0} and \ang{180.0}.
The low panel of \autoref{Stick-Spectrum} includes the possible combination bands,
 i.e.\ those with intensity above the cut off threshold of \SI{0.001}{au}.
The major combination bands are again those associate with the \ce{O-O} and 
 \ce{CH2} wagging modes. 
 
\begin{figure}[htb!]
    \caption{Simulated photoelectron spectrum of \ce{CH2OO-} with the
            stick spectrum shown in \autoref{Stick-Spectrum} convoluted with 
            Gaussian response function (fwhm = \SI{2}{m\electronvolt}).}
    \centering
    \includegraphics[width=1.0\linewidth]{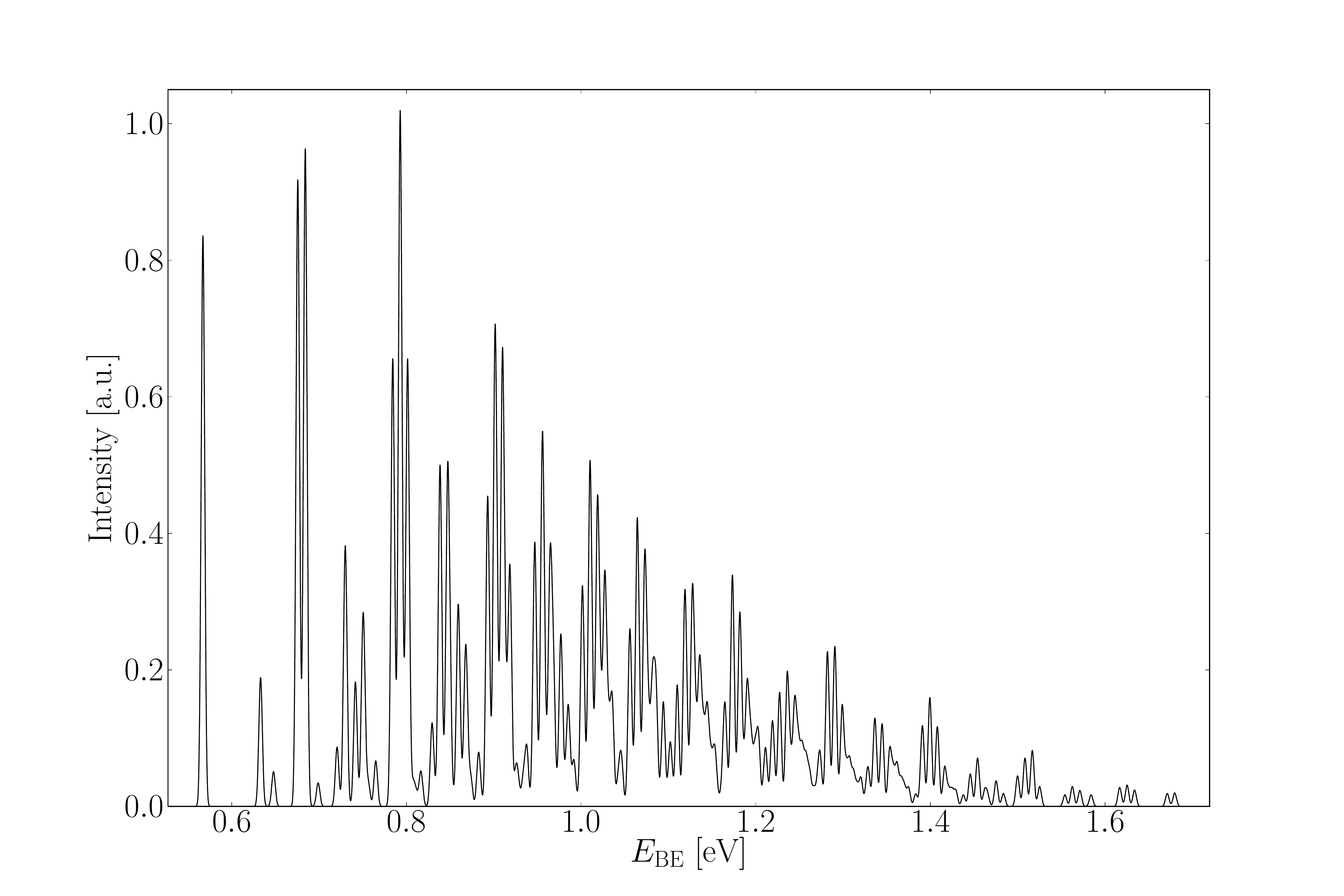}
    \label{Gaussian-Spectrum}
\end{figure}

Again, a full list of line positions including assignments is provided in the
 supplementary material.
To provide a clearer picture of what an experimental spectrum might look like,
 we convoluted the line spectrum with a Gaussian response function whose full 
 width at half maximum was set to \SI{0.002}{\electronvolt} and the resulting
 simulated spectrum is shown in \autoref{Gaussian-Spectrum}.
We believe that this is appropriate as this resolution is achievable by the
 state-of-the-art anion photoelectron spectrometers in use today.

As a final note, when considering the mass selected photoelectron spectrum of
 the \ce{CH2OO-} anion one cannot discount the possibility that a van der Waals
 complex of the form \ce{O-}$\cdots$\ce{CH2O} may also be synthesised in 
 addition to the covalently bound \ce{CH2OO-} anion.
This possibility needs to be taken into account, however, fortuitously the
 ionisation energy of the \ce{O-} anion is a great deal larger than for 
 \ce{CH2OO-}, with the most recent determination of the electron affinity of 
 the \ce{O} neutral being 
 \SI{1.439157}{\electronvolt}.\supercite{joiner.a.11.high-resolution.article}              
The  \ce{O-}$\cdots$\ce{CH2O} species should be essentially a perturbed \ce{O-}
 anion, due to the weak interaction in a van Der Waals species, and in addition
 the electron binding energy of the anion generally increases upon formation of
 a van der Waals complex. 
Therefore, the Criegee anion can be preferentially targeted by selecting a 
 photodetachment photon energy below \SI{1.4}{\electronvolt}.
\section{Summary}

In summary, we have shown that the \ce{CH2OO-} anion with an analogous structure
 to the neutral Criegee intermediate exists.
The anion is stable with respect to dissociation of the \ce{O-O} bond, with a 
 barrier of \SI{16.5}{\kilo\joule\per\mole}, and the electron is bound by 
 \SI{0.567}{\electronvolt}. 
The major geometric differences between the anion and neutral are in the 
 increased \ce{O-O} bond length, and the movement of the terminal hydrogen atoms
 into the plane formed by the \ce{C-O-O} atoms.
Progressions involving both of these modes are prominent in the predicted
 photoelectron spectrum.

\section*{Acknowledgements}
We gratefully acknowledge funding (to A.K.) from the Australian Research Council
 (Discovery Project Grant: DP\num{110102336}) and the generous allocation of
 computing time from the National Computational Infrastructure (NCI) National
 Facility.

\printbibliography

\renewcommand{\appendixname}{Supporting Information}
\setcounter{table}{0}
\renewcommand{\thetable}{S\arabic{table}} 
\appendix
\newpage
\section*{Supporting Information}

\begin{table}[htb!]
\caption{CCSD(T) Optimised Geometries (Cartesian Coordinates in \si{\angstrom}).}
    \label{App: optimised geometries}
    \centering\small
    \begin{tabular}{lSSS}
        \toprule
        \multicolumn{4}{c}{\ce{CH2OO} optimised at CCSD(T)/A$'$VQZ} \\
			& {$x$} & {$y$} & {$z$}							\\ \midrule
		C 	&  1.0747909542	& -0.2065759282 & 0.0000000000  \\
		H 	&  1.9796772459	&  0.3821861731 & 0.0000000000  \\
		O 	& -0.0015157933	&  0.4682056958 & 0.0000000000  \\
		O 	& -1.1650176997 & -0.2029330427 & 0.0000000000  \\
		H 	&  1.0103082928	& -1.2870878981 & 0.0000000000  \\ \midrule
		\multicolumn{4}{c}{\ce{CH2OO-} optimised at CCSD(T)/A$'$VQZ} \\
		C 	&  1.1015331917 & -0.2291495725 & -0.0630600844 \\
		H 	&  2.0160252651 &  0.2927001416 &  0.2177572956 \\
		O 	&  0.0151665961 &  0.5404252237 &  0.0185664201 \\
		O 	& -1.2023201269 & -0.2471468315 & -0.0116962255 \\
		H 	&  0.9654460737 & -1.2824049612 &  0.1624845940 \\ \midrule
		\multicolumn{4}{c}{\ce{CH2OO} optimised at CCSD(T)/A$'$VTZ} \\
		C 	&  1.0768186764 & -0.2070358490 & 0.0000000000  \\
		H 	&  1.9829601118 &  0.3819797961 & 0.0000000000  \\
		O 	& -0.0026186732 &  0.4716100464 & 0.0000000000  \\
		O 	& -1.1712915354 & -0.2041291422 & 0.0000000000  \\
		H 	&  1.0123744203 & -1.2886298513 & 0.0000000000  \\ \midrule
		\multicolumn{4}{c}{\ce{CH2OO-} optimised at CCSD(T)/A$'$VTZ} \\
		C 	&  1.1036992353 & -0.2295959950 & -0.0672636934 \\
		H 	&  2.0183679377 &  0.2917114564 &  0.2196384410 \\
		O 	&  0.0149657098 &  0.5440386206 &  0.0192806220 \\
		O 	& -1.2088036531 & -0.2487240663 & -0.0117291828 \\
		H 	&  0.9676217702 & -1.2830060157 &  0.1641258132 \\ \midrule
		\multicolumn{4}{c}{\ce{[CH2O{\cdots}O]-} optimised at CCSD(T)/A$'$VTZ} \\
		C   &  1.0743808452 & -0.2327031566 & -0.0274002080 \\
		H   &  1.0600867860 & -1.0855298180 & -0.7102234999 \\
		O   &  0.1018213019 &  0.6509445985 & -0.1165934275 \\
		O   & -1.1654217970 & -0.2829429835 &  0.1011911287 \\
		H   &  1.6181098638 & -0.3354896404 &  0.9196730067 \\
		\bottomrule
    \end{tabular}
\end{table}

\begin{table}[htb!]
\caption{Comparison between W2-F12 (this work) and W1$^a$ CCSD(T)/CBS total
atomisation energies for the Criegee intermediate (in \si{\kilo\joule\per\mole}).}
    \label{App: Comparrison W2-F12,W1a}
    \centering\small
    \begin{tabular}{@{}r@{}lSSSSS}
        \toprule
		&			& {CCSD(T)/CBS} & {CV+SR} 	& {SO} & {ZPVE} & {TAE(0)}\\ \midrule
		& W2-F12  	& 1599.9 		& 3.6 		& -2.2 & 80.7 	& 1520.9  \\
		& W1 		& 1598.7 		& 3.7 		& -2.2 & 80.1 	& 1519.1  \\
		\bottomrule
		\parbox[t][][t]{2mm}{$a$} & \multicolumn{6}{l}{\parbox[t][][t]{0.8\linewidth}{Taken from M.T. Nguyen, T.L. Nguyen, V.T. Ngan, H.M.T. Nguyen, Chem. Phys. Lett., 448 (2007) 183.}}
    \end{tabular}
\end{table}

\begin{table}
\caption{Diagnostics for importance of nondynamical correlation.}
    \label{App: Diagnostics}
    \centering\small
    \begin{tabular}{@{}l@{}llS[table-column-width = 3 cm]S[table-column-width = 3 cm] S[table-column-width = 3 cm]}
        \toprule
        &	& 	& {\ce{CH2OO}} & {\ce{CH2OO-}} & {\ce{[CH2O{\cdots}O]-}}\\ \midrule
        & TAE[SCF]$^a$ 		& [\si{\percent}] &  54.4 &  54.6 &  53.2 \\
		& TAE$_e$[(T)]$^b$   	& [\si{\percent}] &   4.6 &   3.8 &   4.2 \\
		& T1$^c$		    	& 				  & 0.044 & 0.034 & 0.045 \\
		& D1$^c$		    	& 				  & 0.176 & 0.122 & 0.130 \\ \bottomrule
		\parbox[t][][t]{2mm}{$a$} & \multicolumn{5}{l}{\parbox[t][][t]{0.85\linewidth}{Percentages of the total atomisation energy accounted by the SCF component relative to nonrelativistic, clamped-nuclei, valence CCSDT(Q) values from W3-F12 theory.}}\\
		\parbox[t][][t]{2mm}{$b$} & \multicolumn{5}{l}{\parbox[t][][t]{0.85\linewidth}{Percentages of the total atomisation energy accounted by the (T) component relative to nonrelativistic, clamped-nuclei, valence CCSDT(Q) values from W3-F12 theory.}}\\
		\parbox[t][][t]{2mm}{$c$} & \multicolumn{5}{l}{\parbox[t][][t]{0.85\linewidth}{From a CCSD(T)/A$'$VTZ calculation.}}

	\end{tabular}
\end{table}

\begin{table}[htb!]
\caption{Orbital occupancies from NBO 6.0 calculations for anion and neutral \ce{CH2OO}.}
    \label{App: NBO}
    \centering\small
    \begin{tabular}{lSS}
    \toprule
    	& \ce{CH2OO}	& \ce{CH2OO-} \\
    \midrule
	\ce{C-O} bonding				& 3.9959			& 2.9960 \\
	\ce{O-O} bonding				& 1.9913			& 1.9873 \\
	\ce{C-O} anti-bonding			& 0.1316			& 0.0180 \\
	\ce{O-O} anti-bonding			& 0.0085			& 0.0088 \\
	\ce{C} lone pair				& 0.0000			& 0.9933 \\
	\ce{O} lone pair				& 1.9780			& 2.9706 \\
	\ce{O} lone pair (terminal)		& 5.8510			& 5.9699 \\
	\ce{C-H} bonding				& 1.9891/1.9955		& 1.9899/1.9956 \\
	\ce{C-H} anti-bonding			& 0.0115/0.0153		& 0.0127/0.0166 \\
	\bottomrule
	\end{tabular}
\end{table}

\newpage

\begin{center}
	\footnotesize
	\begin{longtable}{S S[table-format = 1.6e1, table-column-width = 3cm] S[table-format = 1.6e1, table-column-width = 3cm] l}
		\caption{ezSpectrum 3.0 simulations from CCSD(T)/A$'$VQZ geometries,
     	harmonic frequencies and normal mode vectors. 
     	Data pertain to application of the
     	Duschinsky formalism. Simulation temperature of \SI{10}{K}, 
     	intensity threshold of \SI{1.0e-3}{au}, maximum number of quanta in 
     	excited state $v$ = \num{10}.}
     	\label{App: Duschinski Data}
		\\\toprule
		{$E_{\mathrm{BH}}$ [\si{\electronvolt}]} & {Intensity [au]} & {FCF [au]} & Transition
		\\\midrule[\heavyrulewidth]
		\endfirsthead

		\caption[]{\tablename\ \thetable\ -- \textit{Continued from previous page}}
		\\\toprule
		{$E_{\mathrm{BH}}$ [\si{\electronvolt}]} & {Intensity [au]} & {FCF [au]} & Transition
		\\\midrule[\heavyrulewidth]
		\endhead
		
		\bottomrule
		\multicolumn{4}{r}{\textit{Continued on next page}}
		\endfoot

		\bottomrule
		\endlastfoot

		0.5670	&	2.494071e-02	&	1.579263e-01	&	0(0){$\rightarrow$}1(0)	\\
		0.6330	&	5.634408e-03	&	7.506269e-02	&	0(0){$\rightarrow$}1(1v7)	\\
		0.6478	&	1.523071e-03	&	3.902654e-02	&	0(0){$\rightarrow$}1(1v9)	\\
		0.6756	&	2.739918e-02	&	1.655270e-01	&	0(0){$\rightarrow$}1(1v8)	\\
		0.6842	&	2.875963e-02	&	-1.695866e-01	&	0(0){$\rightarrow$}1(1v6)	\\
		0.6989	&	1.029376e-03	&	3.208390e-02	&	0(0){$\rightarrow$}1(2v7)	\\
		0.7206	&	2.595634e-03	&	5.094736e-02	&	0(0){$\rightarrow$}1(1v5)	\\
		0.7285	&	1.042241e-03	&	3.228376e-02	&	0(0){$\rightarrow$}1(2v9)	\\
		0.7301	&	1.063235e-02	&	-1.031133e-01	&	0(0){$\rightarrow$}1(1v4)	\\
		0.7416	&	5.450907e-03	&	7.383026e-02	&	0(0){$\rightarrow$}1(1v7,1v8)	\\
		0.7501	&	6.244429e-03	&	-7.902170e-02	&	0(0){$\rightarrow$}1(1v7,1v6)	\\
		0.7516	&	2.734344e-03	&	5.229096e-02	&	0(0){$\rightarrow$}1(1v3)	\\
		0.7564	&	9.948150e-04	&	3.154069e-02	&	0(0){$\rightarrow$}1(1v9,1v8)	\\
		0.7649	&	1.995556e-03	&	-4.467164e-02	&	0(0){$\rightarrow$}1(1v9,1v6)	\\
		0.7843	&	1.913721e-02	&	1.383373e-01	&	0(0){$\rightarrow$}1(2v8)	\\
		0.7866	&	8.349666e-04	&	2.889579e-02	&	0(0){$\rightarrow$}1(1v7,1v5)	\\
		0.7928	&	2.983042e-02	&	-1.727148e-01	&	0(0){$\rightarrow$}1(1v8,1v6)	\\
		0.7961	&	2.198228e-03	&	-4.688526e-02	&	0(0){$\rightarrow$}1(1v7,1v4)	\\
		0.8014	&	1.948531e-02	&	1.395898e-01	&	0(0){$\rightarrow$}1(2v6)	\\
		0.8076	&	9.224151e-04	&	3.037129e-02	&	0(0){$\rightarrow$}1(2v7,1v8)	\\
		0.8109	&	5.766484e-04	&	-2.401350e-02	&	0(0){$\rightarrow$}1(1v9,1v4)	\\
		0.8161	&	1.114799e-03	&	-3.338860e-02	&	0(0){$\rightarrow$}1(2v7,1v6)	\\
		0.8176	&	5.241314e-04	&	2.289392e-02	&	0(0){$\rightarrow$}1(1v7,1v3)	\\
		0.8292	&	3.204468e-03	&	5.660802e-02	&	0(0){$\rightarrow$}1(1v8,1v5)	\\
		0.8309	&	6.163621e-04	&	-2.482664e-02	&	0(0){$\rightarrow$}1(1v9,1v7,1v6)	\\
		0.8372	&	9.342162e-04	&	3.056495e-02	&	0(0){$\rightarrow$}1(2v9,1v8)	\\
		0.8378	&	3.426701e-03	&	-5.853803e-02	&	0(0){$\rightarrow$}1(1v6,1v5)	\\
		0.8388	&	1.106530e-02	&	-1.051917e-01	&	0(0){$\rightarrow$}1(1v8,1v4)	\\
		0.8457	&	1.269727e-03	&	-3.563322e-02	&	0(0){$\rightarrow$}1(2v9,1v6)	\\
		0.8473	&	1.292793e-02	&	1.137011e-01	&	0(0){$\rightarrow$}1(1v6,1v4)	\\
		0.8502	&	3.430479e-03	&	5.857029e-02	&	0(0){$\rightarrow$}1(1v7,2v8)	\\
		0.8588	&	5.659486e-03	&	-7.522956e-02	&	0(0){$\rightarrow$}1(1v7,1v8,1v6)	\\
		0.8603	&	3.798860e-03	&	6.163489e-02	&	0(0){$\rightarrow$}1(1v8,1v3)	\\
		0.8673	&	4.088714e-03	&	6.394305e-02	&	0(0){$\rightarrow$}1(1v7,2v6)	\\
		0.8688	&	3.480573e-03	&	-5.899638e-02	&	0(0){$\rightarrow$}1(1v6,1v3)	\\
		0.8736	&	1.261825e-03	&	-3.552217e-02	&	0(0){$\rightarrow$}1(1v9,1v8,1v6)	\\
		0.8821	&	1.497083e-03	&	3.869216e-02	&	0(0){$\rightarrow$}1(1v9,2v6)	\\
		0.8837	&	1.066484e-03	&	-3.265707e-02	&	0(0){$\rightarrow$}1(1v5,1v4)	\\
		0.8929	&	1.054001e-02	&	1.026646e-01	&	0(0){$\rightarrow$}1(3v8)	\\
		0.8933	&	2.558704e-03	&	5.058364e-02	&	0(0){$\rightarrow$}1(2v4)	\\
		0.8952	&	9.149745e-04	&	3.024855e-02	&	0(0){$\rightarrow$}1(1v7,1v8,1v5)	\\
		0.9015	&	1.990639e-02	&	-1.410900e-01	&	0(0){$\rightarrow$}1(2v8,1v6)	\\
		0.9037	&	1.045778e-03	&	-3.233848e-02	&	0(0){$\rightarrow$}1(1v7,1v6,1v5)	\\
		0.9047	&	1.989827e-03	&	-4.460747e-02	&	0(0){$\rightarrow$}1(1v7,1v8,1v4)	\\
		0.9100	&	1.922762e-02	&	1.386637e-01	&	0(0){$\rightarrow$}1(1v8,2v6)	\\
		0.9133	&	2.573238e-03	&	5.072710e-02	&	0(0){$\rightarrow$}1(1v7,1v6,1v4)	\\
		0.9148	&	1.247227e-03	&	-3.531610e-02	&	0(0){$\rightarrow$}1(1v4,1v3)	\\
		0.9162	&	5.487538e-04	&	2.342549e-02	&	0(0){$\rightarrow$}1(2v7,2v8)	\\
		0.9186	&	9.991816e-03	&	-9.995907e-02	&	0(0){$\rightarrow$}1(3v6)	\\
		0.9248	&	9.329596e-04	&	-3.054439e-02	&	0(0){$\rightarrow$}1(2v7,1v8,1v6)	\\
		0.9262	&	6.572315e-04	&	2.563653e-02	&	0(0){$\rightarrow$}1(1v7,1v8,1v3)	\\
		0.9281	&	8.023621e-04	&	2.832600e-02	&	0(0){$\rightarrow$}1(1v9,1v6,1v4)	\\
		0.9333	&	7.160158e-04	&	2.675847e-02	&	0(0){$\rightarrow$}1(2v7,2v6)	\\
		0.9348	&	6.457742e-04	&	-2.541209e-02	&	0(0){$\rightarrow$}1(1v7,1v6,1v3)	\\
		0.9379	&	2.438783e-03	&	4.938404e-02	&	0(0){$\rightarrow$}1(2v8,1v5)	\\
		0.9458	&	5.752186e-04	&	2.398371e-02	&	0(0){$\rightarrow$}1(2v9,2v8)	\\
		0.9464	&	3.991340e-03	&	-6.317705e-02	&	0(0){$\rightarrow$}1(1v8,1v6,1v5)	\\
		0.9474	&	7.404273e-03	&	-8.604809e-02	&	0(0){$\rightarrow$}1(2v8,1v4)	\\
		0.9544	&	1.061017e-03	&	-3.257325e-02	&	0(0){$\rightarrow$}1(2v9,1v8,1v6)	\\
		0.9550	&	2.585400e-03	&	5.084683e-02	&	0(0){$\rightarrow$}1(2v6,1v5)	\\
		0.9560	&	1.272149e-02	&	1.127896e-01	&	0(0){$\rightarrow$}1(1v8,1v6,1v4)	\\
		0.9589	&	1.717748e-03	&	4.144572e-02	&	0(0){$\rightarrow$}1(1v7,3v8)	\\
		0.9629	&	9.030295e-04	&	3.005045e-02	&	0(0){$\rightarrow$}1(2v9,2v6)	\\
		0.9645	&	9.150493e-03	&	-9.565821e-02	&	0(0){$\rightarrow$}1(2v6,1v4)	\\
		0.9674	&	3.384912e-03	&	-5.818000e-02	&	0(0){$\rightarrow$}1(1v7,2v8,1v6)	\\
		0.9689	&	3.127578e-03	&	5.592475e-02	&	0(0){$\rightarrow$}1(2v8,1v3)	\\
		0.9760	&	3.501801e-03	&	5.917602e-02	&	0(0){$\rightarrow$}1(1v7,1v8,2v6)	\\
		0.9775	&	4.556307e-03	&	-6.750042e-02	&	0(0){$\rightarrow$}1(1v8,1v6,1v3)	\\
		0.9822	&	5.200151e-04	&	-2.280384e-02	&	0(0){$\rightarrow$}1(1v9,2v8,1v6)	\\
		0.9845	&	2.032836e-03	&	-4.508698e-02	&	0(0){$\rightarrow$}1(1v7,3v6)	\\
		0.9860	&	2.554232e-03	&	5.053941e-02	&	0(0){$\rightarrow$}1(2v6,1v3)	\\
		0.9908	&	9.159001e-04	&	3.026384e-02	&	0(0){$\rightarrow$}1(1v9,1v8,2v6)	\\
		0.9924	&	1.252689e-03	&	-3.539335e-02	&	0(0){$\rightarrow$}1(1v8,1v5,1v4)	\\
		0.9993	&	8.371008e-04	&	-2.893269e-02	&	0(0){$\rightarrow$}1(1v9,3v6)	\\
		1.0009	&	1.484399e-03	&	3.852790e-02	&	0(0){$\rightarrow$}1(1v6,1v5,1v4)	\\
		1.0016	&	4.984658e-03	&	7.060211e-02	&	0(0){$\rightarrow$}1(4v8)	\\
		1.0019	&	2.535624e-03	&	5.035498e-02	&	0(0){$\rightarrow$}1(1v8,2v4)	\\
		1.0039	&	6.308203e-04	&	2.511614e-02	&	0(0){$\rightarrow$}1(1v7,2v8,1v5)	\\
		1.0088	&	5.422650e-04	&	2.328658e-02	&	0(0){$\rightarrow$}1(2v9,1v6,1v4)	\\
		1.0101	&	1.052832e-02	&	-1.026076e-01	&	0(0){$\rightarrow$}1(3v8,1v6)	\\
		1.0104	&	3.256416e-03	&	-5.706501e-02	&	0(0){$\rightarrow$}1(1v6,2v4)	\\
		1.0124	&	1.074757e-03	&	-3.278348e-02	&	0(0){$\rightarrow$}1(1v7,1v8,1v6,1v5)	\\
		1.0134	&	1.188439e-03	&	-3.447374e-02	&	0(0){$\rightarrow$}1(1v7,2v8,1v4)	\\
		1.0187	&	1.234426e-02	&	1.111047e-01	&	0(0){$\rightarrow$}1(2v8,2v6)	\\
		1.0209	&	7.548573e-04	&	2.747467e-02	&	0(0){$\rightarrow$}1(1v7,2v6,1v5)	\\
		1.0219	&	2.184878e-03	&	4.674268e-02	&	0(0){$\rightarrow$}1(1v7,1v8,1v6,1v4)	\\
		1.0234	&	1.641748e-03	&	-4.051849e-02	&	0(0){$\rightarrow$}1(1v8,1v4,1v3)	\\
		1.0272	&	9.421348e-03	&	-9.706363e-02	&	0(0){$\rightarrow$}1(1v8,3v6)	\\
		1.0305	&	1.762033e-03	&	-4.197658e-02	&	0(0){$\rightarrow$}1(1v7,2v6,1v4)	\\
		1.0319	&	1.661851e-03	&	4.076580e-02	&	0(0){$\rightarrow$}1(1v6,1v4,1v3)	\\
		1.0334	&	5.269648e-04	&	-2.295571e-02	&	0(0){$\rightarrow$}1(2v7,2v8,1v6)	\\
		1.0357	&	4.274924e-03	&	6.538290e-02	&	0(0){$\rightarrow$}1(4v6)	\\
		1.0419	&	5.648697e-04	&	2.376699e-02	&	0(0){$\rightarrow$}1(2v7,1v8,2v6)	\\
		1.0434	&	7.568890e-04	&	-2.751162e-02	&	0(0){$\rightarrow$}1(1v7,1v8,1v6,1v3)	\\
		1.0453	&	6.327986e-04	&	-2.515549e-02	&	0(0){$\rightarrow$}1(1v9,2v6,1v4)	\\
		1.0465	&	1.443090e-03	&	3.798803e-02	&	0(0){$\rightarrow$}1(3v8,1v5)	\\
		1.0551	&	2.897957e-03	&	-5.383267e-02	&	0(0){$\rightarrow$}1(2v8,1v6,1v5)	\\
		1.0561	&	3.925760e-03	&	-6.265589e-02	&	0(0){$\rightarrow$}1(3v8,1v4)	\\
		1.0572	&	1.457639e-03	&	-3.817903e-02	&	0(0){$\rightarrow$}1(1v9,1v1)	\\
		1.0630	&	6.189798e-04	&	-2.487930e-02	&	0(0){$\rightarrow$}1(2v9,2v8,1v6)	\\
		1.0636	&	2.861335e-03	&	5.349144e-02	&	0(0){$\rightarrow$}1(1v8,2v6,1v5)	\\
		1.0646	&	8.147827e-03	&	9.026531e-02	&	0(0){$\rightarrow$}1(2v8,1v6,1v4)	\\
		1.0649	&	1.205411e-03	&	3.471903e-02	&	0(0){$\rightarrow$}1(1v8,1v2)	\\
		1.0675	&	7.429983e-04	&	2.725800e-02	&	0(0){$\rightarrow$}1(1v7,4v8)	\\
		1.0715	&	7.100090e-04	&	2.664599e-02	&	0(0){$\rightarrow$}1(2v9,1v8,2v6)	\\
		1.0722	&	1.452298e-03	&	-3.810903e-02	&	0(0){$\rightarrow$}1(3v6,1v5)	\\
		1.0731	&	8.572458e-03	&	-9.258757e-02	&	0(0){$\rightarrow$}1(1v8,2v6,1v4)	\\
		1.0761	&	1.619512e-03	&	-4.024317e-02	&	0(0){$\rightarrow$}1(1v7,3v8,1v6)	\\
		1.0764	&	5.979348e-04	&	-2.445271e-02	&	0(0){$\rightarrow$}1(1v7,1v6,2v4)	\\
		1.0776	&	1.967069e-03	&	4.435165e-02	&	0(0){$\rightarrow$}1(3v8,1v3)	\\
		1.0817	&	4.874711e-03	&	6.981913e-02	&	0(0){$\rightarrow$}1(3v6,1v4)	\\
		1.0846	&	2.006310e-03	&	4.479185e-02	&	0(0){$\rightarrow$}1(1v7,2v8,2v6)	\\
		1.0861	&	3.572140e-03	&	-5.976738e-02	&	0(0){$\rightarrow$}1(2v8,1v6,1v3)	\\
		1.0932	&	1.653251e-03	&	-4.066019e-02	&	0(0){$\rightarrow$}1(1v7,1v8,3v6)	\\
		1.0946	&	3.174987e-03	&	5.634702e-02	&	0(0){$\rightarrow$}1(1v8,2v6,1v3)	\\
		1.1010	&	9.150417e-04	&	-3.024966e-02	&	0(0){$\rightarrow$}1(2v8,1v5,1v4)	\\
		1.1017	&	8.452092e-04	&	2.907248e-02	&	0(0){$\rightarrow$}1(1v7,4v6)	\\
		1.1032	&	1.402676e-03	&	-3.745231e-02	&	0(0){$\rightarrow$}1(3v6,1v3)	\\
		1.1096	&	1.646926e-03	&	4.058233e-02	&	0(0){$\rightarrow$}1(1v8,1v6,1v5,1v4)	\\
		1.1102	&	2.113242e-03	&	4.597001e-02	&	0(0){$\rightarrow$}1(5v8)	\\
		1.1105	&	1.633327e-03	&	4.041445e-02	&	0(0){$\rightarrow$}1(2v8,2v4)	\\
		1.1181	&	1.171091e-03	&	-3.422121e-02	&	0(0){$\rightarrow$}1(2v6,1v5,1v4)	\\
		1.1188	&	4.798387e-03	&	-6.927039e-02	&	0(0){$\rightarrow$}1(4v8,1v6)	\\
		1.1191	&	3.053950e-03	&	-5.526255e-02	&	0(0){$\rightarrow$}1(1v8,1v6,2v4)	\\
		1.1210	&	7.039515e-04	&	-2.653208e-02	&	0(0){$\rightarrow$}1(1v7,2v8,1v6,1v5)	\\
		1.1220	&	5.677181e-04	&	-2.382684e-02	&	0(0){$\rightarrow$}1(1v7,3v8,1v4)	\\
		1.1273	&	6.310346e-03	&	7.943768e-02	&	0(0){$\rightarrow$}1(3v8,2v6)	\\
		1.1276	&	2.394883e-03	&	4.893754e-02	&	0(0){$\rightarrow$}1(2v6,2v4)	\\
		1.1296	&	7.332451e-04	&	2.707850e-02	&	0(0){$\rightarrow$}1(1v7,1v8,2v6,1v5)	\\
		1.1306	&	1.242485e-03	&	3.524890e-02	&	0(0){$\rightarrow$}1(1v7,2v8,1v6,1v4)	\\
		1.1311	&	5.561488e-04	&	-2.358281e-02	&	0(0){$\rightarrow$}1(1v8,1v6,1v5,1v3)	\\
		1.1320	&	1.292486e-03	&	-3.595116e-02	&	0(0){$\rightarrow$}1(2v8,1v4,1v3)	\\
		1.1358	&	5.842437e-03	&	-7.643583e-02	&	0(0){$\rightarrow$}1(2v8,3v6)	\\
		1.1391	&	1.414344e-03	&	-3.760776e-02	&	0(0){$\rightarrow$}1(1v7,1v8,2v6,1v4)	\\
		1.1406	&	2.064801e-03	&	4.544008e-02	&	0(0){$\rightarrow$}1(1v8,1v6,1v4,1v3)	\\
		1.1444	&	3.863561e-03	&	6.215755e-02	&	0(0){$\rightarrow$}1(1v8,4v6)	\\
		1.1476	&	9.107960e-04	&	3.017940e-02	&	0(0){$\rightarrow$}1(1v7,3v6,1v4)	\\
		1.1491	&	1.267901e-03	&	-3.560760e-02	&	0(0){$\rightarrow$}1(2v6,1v4,1v3)	\\
		1.1521	&	5.367175e-04	&	-2.316716e-02	&	0(0){$\rightarrow$}1(1v7,2v8,1v6,1v3)	\\
		1.1529	&	1.604950e-03	&	-4.006182e-02	&	0(0){$\rightarrow$}1(5v6)	\\
		1.1552	&	7.262036e-04	&	2.694817e-02	&	0(0){$\rightarrow$}1(4v8,1v5)	\\
		1.1606	&	5.086544e-04	&	2.255337e-02	&	0(0){$\rightarrow$}1(1v7,1v8,2v6,1v3)	\\
		1.1637	&	1.644763e-03	&	-4.055568e-02	&	0(0){$\rightarrow$}1(3v8,1v6,1v5)	\\
		1.1647	&	1.793326e-03	&	-4.234768e-02	&	0(0){$\rightarrow$}1(4v8,1v4)	\\
		1.1658	&	1.469467e-03	&	-3.833363e-02	&	0(0){$\rightarrow$}1(1v9,1v8,1v1)	\\
		1.1723	&	1.994930e-03	&	4.466464e-02	&	0(0){$\rightarrow$}1(2v8,2v6,1v5)	\\
		1.1732	&	4.155773e-03	&	6.446528e-02	&	0(0){$\rightarrow$}1(3v8,1v6,1v4)	\\
		1.1736	&	6.020699e-04	&	2.453711e-02	&	0(0){$\rightarrow$}1(1v6,3v4)	\\
		1.1736	&	2.123509e-03	&	4.608154e-02	&	0(0){$\rightarrow$}1(2v8,1v2)	\\
		1.1743	&	1.708268e-03	&	4.133120e-02	&	0(0){$\rightarrow$}1(1v9,1v6,1v1)	\\
		1.1808	&	1.534052e-03	&	-3.916698e-02	&	0(0){$\rightarrow$}1(1v8,3v6,1v5)	\\
		1.1818	&	5.288105e-03	&	-7.271935e-02	&	0(0){$\rightarrow$}1(2v8,2v6,1v4)	\\
		1.1821	&	1.547056e-03	&	-3.933263e-02	&	0(0){$\rightarrow$}1(1v8,1v6,1v2)	\\
		1.1847	&	6.718852e-04	&	-2.592075e-02	&	0(0){$\rightarrow$}1(1v7,4v8,1v6)	\\
		1.1862	&	1.040294e-03	&	3.225359e-02	&	0(0){$\rightarrow$}1(4v8,1v3)	\\
		1.1893	&	6.728731e-04	&	2.593980e-02	&	0(0){$\rightarrow$}1(4v6,1v5)	\\
		1.1903	&	4.365628e-03	&	6.607290e-02	&	0(0){$\rightarrow$}1(1v8,3v6,1v4)	\\
		1.1933	&	9.240582e-04	&	3.039832e-02	&	0(0){$\rightarrow$}1(1v7,3v8,2v6)	\\
		1.1947	&	2.151855e-03	&	-4.638809e-02	&	0(0){$\rightarrow$}1(3v8,1v6,1v3)	\\
		1.1989	&	2.158090e-03	&	-4.645525e-02	&	0(0){$\rightarrow$}1(4v6,1v4)	\\
		1.2018	&	9.114148e-04	&	-3.018965e-02	&	0(0){$\rightarrow$}1(1v7,2v8,3v6)	\\
		1.2033	&	2.387240e-03	&	4.885939e-02	&	0(0){$\rightarrow$}1(2v8,2v6,1v3)	\\
		1.2097	&	5.219449e-04	&	-2.284611e-02	&	0(0){$\rightarrow$}1(3v8,1v5,1v4)	\\
		1.2103	&	6.549313e-04	&	2.559163e-02	&	0(0){$\rightarrow$}1(1v7,1v8,4v6)	\\
		1.2118	&	1.663499e-03	&	-4.078602e-02	&	0(0){$\rightarrow$}1(1v8,3v6,1v3)	\\
		1.2182	&	1.149491e-03	&	3.390415e-02	&	0(0){$\rightarrow$}1(2v8,1v6,1v5,1v4)	\\
		1.2189	&	8.241443e-04	&	2.870791e-02	&	0(0){$\rightarrow$}1(6v8)	\\
		1.2192	&	8.374324e-04	&	2.893843e-02	&	0(0){$\rightarrow$}1(3v8,2v4)	\\
		1.2203	&	6.398558e-04	&	2.529537e-02	&	0(0){$\rightarrow$}1(1v9,1v4,1v1)	\\
		1.2204	&	6.375547e-04	&	2.524985e-02	&	0(0){$\rightarrow$}1(4v6,1v3)	\\
		1.2267	&	1.235274e-03	&	-3.514646e-02	&	0(0){$\rightarrow$}1(1v8,2v6,1v5,1v4)	\\
		1.2274	&	1.965349e-03	&	-4.433226e-02	&	0(0){$\rightarrow$}1(5v8,1v6)	\\
		1.2277	&	1.885351e-03	&	-4.342063e-02	&	0(0){$\rightarrow$}1(2v8,1v6,2v4)	\\
		1.2353	&	6.841581e-04	&	2.615642e-02	&	0(0){$\rightarrow$}1(3v6,1v5,1v4)	\\
		1.2359	&	2.788909e-03	&	5.281012e-02	&	0(0){$\rightarrow$}1(4v8,2v6)	\\
		1.2363	&	2.139057e-03	&	4.624994e-02	&	0(0){$\rightarrow$}1(1v8,2v6,2v4)	\\
		1.2392	&	5.681265e-04	&	2.383540e-02	&	0(0){$\rightarrow$}1(1v7,3v8,1v6,1v4)	\\
		1.2395	&	5.974229e-04	&	2.444224e-02	&	0(0){$\rightarrow$}1(1v7,2v8,1v2)	\\
		1.2407	&	7.813396e-04	&	-2.795245e-02	&	0(0){$\rightarrow$}1(3v8,1v4,1v3)	\\
		1.2445	&	2.897225e-03	&	-5.382588e-02	&	0(0){$\rightarrow$}1(3v8,3v6)	\\
		1.2448	&	1.319568e-03	&	-3.632586e-02	&	0(0){$\rightarrow$}1(3v6,2v4)	\\
		1.2477	&	7.712887e-04	&	-2.777208e-02	&	0(0){$\rightarrow$}1(1v7,2v8,2v6,1v4)	\\
		1.2481	&	5.321856e-04	&	-2.306915e-02	&	0(0){$\rightarrow$}1(1v7,1v8,1v6,1v2)	\\
		1.2492	&	1.550973e-03	&	3.938240e-02	&	0(0){$\rightarrow$}1(2v8,1v6,1v4,1v3)	\\
		1.2530	&	2.320834e-03	&	4.817503e-02	&	0(0){$\rightarrow$}1(2v8,4v6)	\\
		1.2563	&	6.942741e-04	&	2.634908e-02	&	0(0){$\rightarrow$}1(1v7,1v8,3v6,1v4)	\\
		1.2578	&	1.497392e-03	&	-3.869615e-02	&	0(0){$\rightarrow$}1(1v8,2v6,1v4,1v3)	\\
		1.2616	&	1.393528e-03	&	-3.732999e-02	&	0(0){$\rightarrow$}1(1v8,5v6)	\\
		1.2663	&	7.206741e-04	&	2.684537e-02	&	0(0){$\rightarrow$}1(3v6,1v4,1v3)	\\
		1.2701	&	5.450301e-04	&	2.334588e-02	&	0(0){$\rightarrow$}1(6v6)	\\
		1.2724	&	7.968214e-04	&	-2.822802e-02	&	0(0){$\rightarrow$}1(4v8,1v6,1v5)	\\
		1.2733	&	7.361233e-04	&	-2.713159e-02	&	0(0){$\rightarrow$}1(5v8,1v4)	\\
		1.2744	&	9.953026e-04	&	-3.154842e-02	&	0(0){$\rightarrow$}1(1v9,2v8,1v1)	\\
		1.2809	&	1.092626e-03	&	3.305490e-02	&	0(0){$\rightarrow$}1(3v8,2v6,1v5)	\\
		1.2819	&	1.832699e-03	&	4.281003e-02	&	0(0){$\rightarrow$}1(4v8,1v6,1v4)	\\
		1.2822	&	5.398013e-04	&	2.323362e-02	&	0(0){$\rightarrow$}1(1v8,1v6,3v4)	\\
		1.2822	&	2.043919e-03	&	4.520972e-02	&	0(0){$\rightarrow$}1(3v8,1v2)	\\
		1.2830	&	1.606832e-03	&	4.008531e-02	&	0(0){$\rightarrow$}1(1v9,1v8,1v6,1v1)	\\
		1.2894	&	1.031252e-03	&	-3.211312e-02	&	0(0){$\rightarrow$}1(2v8,3v6,1v5)	\\
		1.2904	&	2.609689e-03	&	-5.108512e-02	&	0(0){$\rightarrow$}1(3v8,2v6,1v4)	\\
		1.2908	&	2.484412e-03	&	-4.984388e-02	&	0(0){$\rightarrow$}1(2v8,1v6,1v2)	\\
		1.2915	&	1.175673e-03	&	-3.428808e-02	&	0(0){$\rightarrow$}1(1v9,2v6,1v1)	\\
		1.2980	&	6.805475e-04	&	2.608731e-02	&	0(0){$\rightarrow$}1(1v8,4v6,1v5)	\\
		1.2990	&	2.603466e-03	&	5.102417e-02	&	0(0){$\rightarrow$}1(2v8,3v6,1v4)	\\
		1.2993	&	1.135841e-03	&	3.370224e-02	&	0(0){$\rightarrow$}1(1v8,2v6,1v2)	\\
		1.3034	&	1.094247e-03	&	-3.307940e-02	&	0(0){$\rightarrow$}1(4v8,1v6,1v3)	\\
		1.3037	&	5.224871e-04	&	-2.285798e-02	&	0(0){$\rightarrow$}1(1v8,1v6,2v4,1v3)	\\
		1.3075	&	1.852977e-03	&	-4.304622e-02	&	0(0){$\rightarrow$}1(1v8,4v6,1v4)	\\
		1.3119	&	1.386489e-03	&	3.723559e-02	&	0(0){$\rightarrow$}1(3v8,2v6,1v3)	\\
		1.3161	&	8.358369e-04	&	2.891084e-02	&	0(0){$\rightarrow$}1(5v6,1v4)	\\
		1.3205	&	1.204721e-03	&	-3.470910e-02	&	0(0){$\rightarrow$}1(2v8,3v6,1v3)	\\
		1.3268	&	6.298888e-04	&	2.509758e-02	&	0(0){$\rightarrow$}1(3v8,1v6,1v5,1v4)	\\
		1.3289	&	6.099998e-04	&	2.469817e-02	&	0(0){$\rightarrow$}1(1v9,1v8,1v4,1v1)	\\
		1.3290	&	7.238332e-04	&	2.690415e-02	&	0(0){$\rightarrow$}1(1v8,4v6,1v3)	\\
		1.3354	&	8.287162e-04	&	-2.878743e-02	&	0(0){$\rightarrow$}1(2v8,2v6,1v5,1v4)	\\
		1.3361	&	7.419347e-04	&	-2.723848e-02	&	0(0){$\rightarrow$}1(6v8,1v6)	\\
		1.3364	&	9.310346e-04	&	-3.051286e-02	&	0(0){$\rightarrow$}1(3v8,1v6,2v4)	\\
		1.3367	&	7.939581e-04	&	-2.817726e-02	&	0(0){$\rightarrow$}1(2v8,1v4,1v2)	\\
		1.3375	&	7.902473e-04	&	-2.811134e-02	&	0(0){$\rightarrow$}1(1v9,1v6,1v4,1v1)	\\
		1.3439	&	6.889895e-04	&	2.624861e-02	&	0(0){$\rightarrow$}1(1v8,3v6,1v5,1v4)	\\
		1.3446	&	1.110272e-03	&	3.332074e-02	&	0(0){$\rightarrow$}1(5v8,2v6)	\\
		1.3449	&	1.272892e-03	&	3.567761e-02	&	0(0){$\rightarrow$}1(2v8,2v6,2v4)	\\
		1.3452	&	6.186973e-04	&	2.487363e-02	&	0(0){$\rightarrow$}1(1v8,1v6,1v4,1v2)	\\
		1.3531	&	1.245923e-03	&	-3.529763e-02	&	0(0){$\rightarrow$}1(4v8,3v6)	\\
		1.3535	&	1.126893e-03	&	-3.356923e-02	&	0(0){$\rightarrow$}1(1v8,3v6,2v4)	\\
		1.3567	&	6.578640e-04	&	-2.564886e-02	&	0(0){$\rightarrow$}1(1v7,2v8,1v6,1v2)	\\
		1.3579	&	8.997283e-04	&	2.999547e-02	&	0(0){$\rightarrow$}1(3v8,1v6,1v4,1v3)	\\
		1.3617	&	1.119296e-03	&	3.345588e-02	&	0(0){$\rightarrow$}1(3v8,4v6)	\\
		1.3620	&	6.022243e-04	&	2.454026e-02	&	0(0){$\rightarrow$}1(4v6,2v4)	\\
		1.3664	&	1.080050e-03	&	-3.286411e-02	&	0(0){$\rightarrow$}1(2v8,2v6,1v4,1v3)	\\
		1.3702	&	8.126162e-04	&	-2.850642e-02	&	0(0){$\rightarrow$}1(2v8,5v6)	\\
		1.3750	&	8.125901e-04	&	2.850597e-02	&	0(0){$\rightarrow$}1(1v8,3v6,1v4,1v3)	\\
		1.3831	&	5.454356e-04	&	-2.335456e-02	&	0(0){$\rightarrow$}1(1v9,3v8,1v1)	\\
		1.3896	&	5.125597e-04	&	2.263978e-02	&	0(0){$\rightarrow$}1(4v8,2v6,1v5)	\\
		1.3905	&	7.280548e-04	&	2.698249e-02	&	0(0){$\rightarrow$}1(5v8,1v6,1v4)	\\
		1.3909	&	1.434008e-03	&	3.786829e-02	&	0(0){$\rightarrow$}1(4v8,1v2)	\\
		1.3916	&	1.032269e-03	&	3.212894e-02	&	0(0){$\rightarrow$}1(1v9,2v8,1v6,1v1)	\\
		1.3981	&	5.470989e-04	&	-2.339014e-02	&	0(0){$\rightarrow$}1(3v8,3v6,1v5)	\\
		1.3991	&	1.117192e-03	&	-3.342442e-02	&	0(0){$\rightarrow$}1(4v8,2v6,1v4)	\\
		1.3994	&	2.249823e-03	&	-4.743230e-02	&	0(0){$\rightarrow$}1(3v8,1v6,1v2)	\\
		1.4002	&	1.041178e-03	&	-3.226728e-02	&	0(0){$\rightarrow$}1(1v9,1v8,2v6,1v1)	\\
		1.4076	&	1.247339e-03	&	3.531768e-02	&	0(0){$\rightarrow$}1(3v8,3v6,1v4)	\\
		1.4079	&	1.692415e-03	&	4.113898e-02	&	0(0){$\rightarrow$}1(2v8,2v6,1v2)	\\
		1.4087	&	6.121766e-04	&	2.474220e-02	&	0(0){$\rightarrow$}1(1v9,3v6,1v1)	\\
		1.4162	&	1.071148e-03	&	-3.272840e-02	&	0(0){$\rightarrow$}1(2v8,4v6,1v4)	\\
		1.4165	&	6.214929e-04	&	-2.492976e-02	&	0(0){$\rightarrow$}1(1v8,3v6,1v2)	\\
		1.4206	&	6.821899e-04	&	2.611877e-02	&	0(0){$\rightarrow$}1(4v8,2v6,1v3)	\\
		1.4247	&	6.895717e-04	&	2.625970e-02	&	0(0){$\rightarrow$}1(1v8,5v6,1v4)	\\
		1.4291	&	6.772435e-04	&	-2.602390e-02	&	0(0){$\rightarrow$}1(3v8,3v6,1v3)	\\
		1.4377	&	5.064710e-04	&	2.250491e-02	&	0(0){$\rightarrow$}1(2v8,4v6,1v3)	\\
		1.4453	&	7.457520e-04	&	-2.730846e-02	&	0(0){$\rightarrow$}1(3v8,1v4,1v2)	\\
		1.4461	&	7.040460e-04	&	-2.653386e-02	&	0(0){$\rightarrow$}1(1v9,1v8,1v6,1v4,1v1)	\\
		1.4536	&	6.086454e-04	&	2.467074e-02	&	0(0){$\rightarrow$}1(3v8,2v6,2v4)	\\
		1.4539	&	9.835989e-04	&	3.136238e-02	&	0(0){$\rightarrow$}1(2v8,1v6,1v4,1v2)	\\
		1.4547	&	5.679067e-04	&	2.383079e-02	&	0(0){$\rightarrow$}1(1v9,2v6,1v4,1v1)	\\
		1.4621	&	6.486771e-04	&	-2.546914e-02	&	0(0){$\rightarrow$}1(2v8,3v6,2v4)	\\
		1.4654	&	5.037241e-04	&	-2.244380e-02	&	0(0){$\rightarrow$}1(1v7,3v8,1v6,1v2)	\\
		1.4751	&	6.047977e-04	&	-2.459263e-02	&	0(0){$\rightarrow$}1(3v8,2v6,1v4,1v3)	\\
		1.4754	&	5.204542e-04	&	-2.281347e-02	&	0(0){$\rightarrow$}1(2v8,1v6,1v3,1v2)	\\
		1.4836	&	5.650680e-04	&	2.377116e-02	&	0(0){$\rightarrow$}1(2v8,3v6,1v4,1v3)	\\
		1.4995	&	8.235563e-04	&	2.869767e-02	&	0(0){$\rightarrow$}1(5v8,1v2)	\\
		1.5003	&	5.398049e-04	&	2.323370e-02	&	0(0){$\rightarrow$}1(1v9,3v8,1v6,1v1)	\\
		1.5081	&	1.502701e-03	&	-3.876469e-02	&	0(0){$\rightarrow$}1(4v8,1v6,1v2)	\\
		1.5088	&	6.396327e-04	&	-2.529096e-02	&	0(0){$\rightarrow$}1(1v9,2v8,2v6,1v1)	\\
		1.5163	&	5.199924e-04	&	2.280334e-02	&	0(0){$\rightarrow$}1(4v8,3v6,1v4)	\\
		1.5166	&	1.457512e-03	&	3.817737e-02	&	0(0){$\rightarrow$}1(3v8,2v6,1v2)	\\
		1.5174	&	5.131045e-04	&	2.265181e-02	&	0(0){$\rightarrow$}1(1v9,1v8,3v6,1v1)	\\
		1.5251	&	8.681297e-04	&	-2.946404e-02	&	0(0){$\rightarrow$}1(2v8,3v6,1v2)	\\
		1.5540	&	5.087894e-04	&	-2.255636e-02	&	0(0){$\rightarrow$}1(4v8,1v4,1v2)	\\
		1.5625	&	8.698106e-04	&	2.949255e-02	&	0(0){$\rightarrow$}1(3v8,1v6,1v4,1v2)	\\
		1.5711	&	7.024421e-04	&	-2.650362e-02	&	0(0){$\rightarrow$}1(2v8,2v6,1v4,1v2)	\\
		1.5840	&	5.057134e-04	&	-2.248807e-02	&	0(0){$\rightarrow$}1(3v8,1v6,1v3,1v2)	\\
		1.6167	&	8.269810e-04	&	-2.875728e-02	&	0(0){$\rightarrow$}1(5v8,1v6,1v2)	\\
		1.6252	&	9.346853e-04	&	3.057262e-02	&	0(0){$\rightarrow$}1(4v8,2v6,1v2)	\\
		1.6338	&	7.154857e-04	&	-2.674856e-02	&	0(0){$\rightarrow$}1(3v8,3v6,1v2)	\\
		1.6712	&	5.656530e-04	&	2.378346e-02	&	0(0){$\rightarrow$}1(4v8,1v6,1v4,1v2)	\\
		1.6797	&	5.908233e-04	&	-2.430686e-02	&	0(0){$\rightarrow$}1(3v8,2v6,1v4,1v2)	\\
	\end{longtable}
\end{center}

\ \\

\begin{center}
\footnotesize
\begin{longtable}{S S[table-format = 1.6e1, table-column-width = 3cm] S[table-format = 1.6e1, table-column-width = 3cm] l}
    \caption{ezSpectrum 3.0 simulations from CCSD(T)/A$'$VQZ geometries,
     harmonic frequencies and normal mode vectors. 
     Data pertain to application of the Parallel Mode formalism.
     Simulation temperature of \SI{10}{K}, 
     intensity threshold of \SI{1.0e-3}{au}, maximum number of quanta in 
     excited state $v$ = \num{10}.}
	 \label{App: Parallel Data}
		\\\toprule
		{$E_{\mathrm{BH}}$ [\si{\electronvolt}]} & {Intensity [au]} & {FCF [au]} & Transition
		\\\midrule[\heavyrulewidth]
		\endfirsthead

		\caption[]{\tablename\ \thetable\ -- \textit{Continued from previous page}}
		\\\toprule
		{$E_{\mathrm{BH}}$ [\si{\electronvolt}]} & {Intensity [au]} & {FCF [au]} & Transition
		\\\midrule[\heavyrulewidth]
		\endhead
		
		\bottomrule
		\multicolumn{4}{r}{\textit{Continued on next page}}
		\endfoot

		\bottomrule
		\endlastfoot
			0.5670	&	2.631643E-2	&	1.622234E-1	&	0(0){$\rightarrow$}1(0)	\\
			0.6330	&	5.001912E-3	&	7.072420E-2	&	0(0){$\rightarrow$}1(1v7)	\\
			0.6478	&	7.771837E-4	&	2.787802E-2	&	0(0){$\rightarrow$}1(1v9)	\\
			0.6756	&	2.577728E-2	&	1.605530E-1	&	0(0){$\rightarrow$}1(1v8)	\\
			0.6842	&	2.997164E-2	&	-1.731232E-1	&	0(0){$\rightarrow$}1(1v6)	\\
			0.6989	&	8.608812E-4	&	2.934078E-2	&	0(0){$\rightarrow$}1(2v7)	\\
			0.7206	&	3.381422E-3	&	5.814999E-2	&	0(0){$\rightarrow$}1(1v5)	\\
			0.7285	&	1.863135E-3	&	4.316405E-2	&	0(0){$\rightarrow$}1(2v9)	\\
			0.7301	&	1.095256E-2	&	-1.046545E-1	&	0(0){$\rightarrow$}1(1v4)	\\
			0.7416	&	4.899438E-3	&	6.999598E-2	&	0(0){$\rightarrow$}1(1v7,1v8)	\\
			0.7501	&	5.696651E-3	&	-7.547616E-2	&	0(0){$\rightarrow$}1(1v7,1v6)	\\
			0.7516	&	4.662671E-3	&	6.828376E-2	&	0(0){$\rightarrow$}1(1v3)	\\
			0.7564	&	7.612615E-4	&	2.759097E-2	&	0(0){$\rightarrow$}1(1v9,1v8)	\\
			0.7649	&	8.851304E-4	&	-2.975114E-2	&	0(0){$\rightarrow$}1(1v9,1v6)	\\
			0.7843	&	1.787438E-2	&	1.336951E-1	&	0(0){$\rightarrow$}1(2v8)	\\
			0.7866	&	6.427003E-4	&	2.535153E-2	&	0(0){$\rightarrow$}1(1v7,1v5)	\\
			0.7928	&	2.935760E-2	&	-1.713406E-1	&	0(0){$\rightarrow$}1(1v8,1v6)	\\
			0.7961	&	2.081732E-3	&	-4.562600E-2	&	0(0){$\rightarrow$}1(1v7,1v4)	\\
			0.8014	&	2.013154E-2	&	1.418857E-1	&	0(0){$\rightarrow$}1(2v6)	\\
			0.8076	&	8.432443E-4	&	2.903867E-2	&	0(0){$\rightarrow$}1(2v7,1v8)	\\
			0.8161	&	9.804530E-4	&	-3.131219E-2	&	0(0){$\rightarrow$}1(2v7,1v6)	\\
			0.8176	&	8.862250E-4	&	2.976953E-2	&	0(0){$\rightarrow$}1(1v7,1v3)	\\
			0.8292	&	3.312146E-3	&	5.755125E-2	&	0(0){$\rightarrow$}1(1v8,1v5)	\\
			0.8372	&	1.824965E-3	&	4.271960E-2	&	0(0){$\rightarrow$}1(2v9,1v8)	\\
			0.8378	&	3.851083E-3	&	-6.205709E-2	&	0(0){$\rightarrow$}1(1v6,1v5)	\\
			0.8388	&	1.072817E-2	&	-1.035769E-1	&	0(0){$\rightarrow$}1(1v8,1v4)	\\
			0.8457	&	2.121914E-3	&	-4.606424E-2	&	0(0){$\rightarrow$}1(2v9,1v6)	\\
			0.8473	&	1.247381E-2	&	1.116862E-1	&	0(0){$\rightarrow$}1(1v6,1v4)	\\
			0.8502	&	3.397348E-3	&	5.828678E-2	&	0(0){$\rightarrow$}1(1v7,2v8)	\\
			0.8588	&	5.579943E-3	&	-7.469902E-2	&	0(0){$\rightarrow$}1(1v7,1v8,1v6)	\\
			0.8603	&	4.567147E-3	&	6.758067E-2	&	0(0){$\rightarrow$}1(1v8,1v3)	\\
			0.8651	&	5.278709E-4	&	2.297544E-2	&	0(0){$\rightarrow$}1(1v9,2v8)	\\
			0.8673	&	3.826363E-3	&	6.185761E-2	&	0(0){$\rightarrow$}1(1v7,2v6)	\\
			0.8688	&	5.310291E-3	&	-7.287175E-2	&	0(0){$\rightarrow$}1(1v6,1v3)	\\
			0.8736	&	8.669966E-4	&	-2.944481E-2	&	0(0){$\rightarrow$}1(1v9,1v8,1v6)	\\
			0.8821	&	5.945301E-4	&	2.438299E-2	&	0(0){$\rightarrow$}1(1v9,2v6)	\\
			0.8837	&	1.407304E-3	&	-3.751406E-2	&	0(0){$\rightarrow$}1(1v5,1v4)	\\
			0.8916	&	7.754129E-4	&	-2.784624E-2	&	0(0){$\rightarrow$}1(2v9,1v4)	\\
			0.8929	&	1.015601E-2	&	1.007770E-1	&	0(0){$\rightarrow$}1(3v8)	\\
			0.8933	&	2.558786E-3	&	5.058445E-2	&	0(0){$\rightarrow$}1(2v4)	\\
			0.8952	&	6.295332E-4	&	2.509050E-2	&	0(0){$\rightarrow$}1(1v7,1v8,1v5)	\\
			0.9015	&	2.035703E-2	&	-1.426781E-1	&	0(0){$\rightarrow$}1(2v8,1v6)	\\
			0.9037	&	7.319679E-4	&	-2.705491E-2	&	0(0){$\rightarrow$}1(1v7,1v6,1v5)	\\
			0.9047	&	2.039083E-3	&	-4.515621E-2	&	0(0){$\rightarrow$}1(1v7,1v8,1v4)	\\
			0.9052	&	5.991109E-4	&	2.447674E-2	&	0(0){$\rightarrow$}1(1v5,1v3)	\\
			0.9100	&	1.971910E-2	&	1.404247E-1	&	0(0){$\rightarrow$}1(1v8,2v6)	\\
			0.9133	&	2.370873E-3	&	4.869161E-2	&	0(0){$\rightarrow$}1(1v7,1v6,1v4)	\\
			0.9148	&	1.940544E-3	&	-4.405161E-2	&	0(0){$\rightarrow$}1(1v4,1v3)	\\
			0.9162	&	5.847190E-4	&	2.418096E-2	&	0(0){$\rightarrow$}1(2v7,2v8)	\\
			0.9186	&	1.025713E-2	&	-1.012775E-1	&	0(0){$\rightarrow$}1(3v6)	\\
			0.9248	&	9.603663E-4	&	-3.098978E-2	&	0(0){$\rightarrow$}1(2v7,1v8,1v6)	\\
			0.9262	&	8.680688E-4	&	2.946301E-2	&	0(0){$\rightarrow$}1(1v7,1v8,1v3)	\\
			0.9333	&	6.585570E-4	&	2.566237E-2	&	0(0){$\rightarrow$}1(2v7,2v6)	\\
			0.9348	&	1.009317E-3	&	-3.176975E-2	&	0(0){$\rightarrow$}1(1v7,1v6,1v3)	\\
			0.9363	&	5.260983E-4	&	2.293683E-2	&	0(0){$\rightarrow$}1(2v3)	\\
			0.9379	&	2.296695E-3	&	4.792385E-2	&	0(0){$\rightarrow$}1(2v8,1v5)	\\
			0.9458	&	1.265460E-3	&	3.557330E-2	&	0(0){$\rightarrow$}1(2v9,2v8)	\\
			0.9464	&	3.772185E-3	&	-6.141812E-2	&	0(0){$\rightarrow$}1(1v8,1v6,1v5)	\\
			0.9474	&	7.439087E-3	&	-8.625014E-2	&	0(0){$\rightarrow$}1(2v8,1v4)	\\
			0.9544	&	2.078442E-3	&	-4.558994E-2	&	0(0){$\rightarrow$}1(2v9,1v8,1v6)	\\
			0.9550	&	2.586720E-3	&	5.085981E-2	&	0(0){$\rightarrow$}1(2v6,1v5)	\\
			0.9560	&	1.221826E-2	&	1.105362E-1	&	0(0){$\rightarrow$}1(1v8,1v6,1v4)	\\
			0.9563	&	2.390178E-3	&	4.888945E-2	&	0(0){$\rightarrow$}1(1v2)	\\
			0.9589	&	1.930333E-3	&	4.393555E-2	&	0(0){$\rightarrow$}1(1v7,3v8)	\\
			0.9629	&	1.425261E-3	&	3.775263E-2	&	0(0){$\rightarrow$}1(2v9,2v6)	\\
			0.9645	&	8.378490E-3	&	-9.153409E-2	&	0(0){$\rightarrow$}1(2v6,1v4)	\\
			0.9674	&	3.869222E-3	&	-6.220307E-2	&	0(0){$\rightarrow$}1(1v7,2v8,1v6)	\\
			0.9689	&	3.166933E-3	&	5.627551E-2	&	0(0){$\rightarrow$}1(2v8,1v3)	\\
			0.9760	&	3.747972E-3	&	6.122068E-2	&	0(0){$\rightarrow$}1(1v7,1v8,2v6)	\\
			0.9775	&	5.201499E-3	&	-7.212142E-2	&	0(0){$\rightarrow$}1(1v8,1v6,1v3)	\\
			0.9822	&	6.011893E-4	&	-2.451916E-2	&	0(0){$\rightarrow$}1(1v9,2v8,1v6)	\\
			0.9845	&	1.949553E-3	&	-4.415374E-2	&	0(0){$\rightarrow$}1(1v7,3v6)	\\
			0.9860	&	3.566851E-3	&	5.972312E-2	&	0(0){$\rightarrow$}1(2v6,1v3)	\\
			0.9908	&	5.823499E-4	&	2.413193E-2	&	0(0){$\rightarrow$}1(1v9,1v8,2v6)	\\
			0.9924	&	1.378473E-3	&	-3.712779E-2	&	0(0){$\rightarrow$}1(1v8,1v5,1v4)	\\
			1.0003	&	7.595269E-4	&	-2.755952E-2	&	0(0){$\rightarrow$}1(2v9,1v8,1v4)	\\
			1.0009	&	1.602771E-3	&	4.003463E-2	&	0(0){$\rightarrow$}1(1v6,1v5,1v4)	\\
			1.0016	&	5.098724E-3	&	7.140535E-2	&	0(0){$\rightarrow$}1(4v8)	\\
			1.0019	&	2.506364E-3	&	5.006360E-2	&	0(0){$\rightarrow$}1(1v8,2v4)	\\
			1.0088	&	8.831135E-4	&	2.971723E-2	&	0(0){$\rightarrow$}1(2v9,1v6,1v4)	\\
			1.0101	&	1.156662E-2	&	-1.075482E-1	&	0(0){$\rightarrow$}1(3v8,1v6)	\\
			1.0104	&	2.914188E-3	&	-5.398322E-2	&	0(0){$\rightarrow$}1(1v6,2v4)	\\
			1.0124	&	7.169720E-4	&	-2.677633E-2	&	0(0){$\rightarrow$}1(1v7,1v8,1v6,1v5)	\\
			1.0134	&	1.413933E-3	&	-3.760230E-2	&	0(0){$\rightarrow$}1(1v7,2v8,1v4)	\\
			1.0139	&	5.868369E-4	&	2.422472E-2	&	0(0){$\rightarrow$}1(1v8,1v5,1v3)	\\
			1.0187	&	1.367354E-2	&	1.169339E-1	&	0(0){$\rightarrow$}1(2v8,2v6)	\\
			1.0219	&	2.322301E-3	&	4.819026E-2	&	0(0){$\rightarrow$}1(1v7,1v8,1v6,1v4)	\\
			1.0224	&	6.823242E-4	&	-2.612134E-2	&	0(0){$\rightarrow$}1(1v6,1v5,1v3)	\\
			1.0234	&	1.900788E-3	&	-4.359803E-2	&	0(0){$\rightarrow$}1(1v8,1v4,1v3)	\\
			1.0272	&	1.004699E-2	&	-1.002347E-1	&	0(0){$\rightarrow$}1(1v8,3v6)	\\
			1.0305	&	1.592484E-3	&	-3.990593E-2	&	0(0){$\rightarrow$}1(1v7,2v6,1v4)	\\
			1.0319	&	2.210075E-3	&	4.701144E-2	&	0(0){$\rightarrow$}1(1v6,1v4,1v3)	\\
			1.0334	&	6.659333E-4	&	-2.580568E-2	&	0(0){$\rightarrow$}1(2v7,2v8,1v6)	\\
			1.0349	&	6.019328E-4	&	2.453432E-2	&	0(0){$\rightarrow$}1(1v7,2v8,1v3)	\\
			1.0357	&	4.367313E-3	&	6.608565E-2	&	0(0){$\rightarrow$}1(4v6)	\\
			1.0419	&	6.450651E-4	&	2.539813E-2	&	0(0){$\rightarrow$}1(2v7,1v8,2v6)	\\
			1.0434	&	9.886389E-4	&	-3.144263E-2	&	0(0){$\rightarrow$}1(1v7,1v8,1v6,1v3)	\\
			1.0449	&	5.153200E-4	&	2.270066E-2	&	0(0){$\rightarrow$}1(1v8,2v3)	\\
			1.0465	&	1.304955E-3	&	3.612415E-2	&	0(0){$\rightarrow$}1(3v8,1v5)	\\
			1.0520	&	6.779445E-4	&	2.603737E-2	&	0(0){$\rightarrow$}1(1v7,2v6,1v3)	\\
			1.0534	&	5.991705E-4	&	-2.447796E-2	&	0(0){$\rightarrow$}1(1v6,2v3)	\\
			1.0545	&	7.190189E-4	&	2.681453E-2	&	0(0){$\rightarrow$}1(2v9,3v8)	\\
			1.0551	&	2.615694E-3	&	-5.114385E-2	&	0(0){$\rightarrow$}1(2v8,1v6,1v5)	\\
			1.0561	&	4.226800E-3	&	-6.501384E-2	&	0(0){$\rightarrow$}1(3v8,1v4)	\\
			1.0630	&	1.441225E-3	&	-3.796347E-2	&	0(0){$\rightarrow$}1(2v9,2v8,1v6)	\\
			1.0636	&	2.533726E-3	&	5.033613E-2	&	0(0){$\rightarrow$}1(1v8,2v6,1v5)	\\
			1.0646	&	8.472336E-3	&	9.204529E-2	&	0(0){$\rightarrow$}1(2v8,1v6,1v4)	\\
			1.0649	&	2.341210E-3	&	4.838606E-2	&	0(0){$\rightarrow$}1(1v8,1v2)	\\
			1.0675	&	9.691047E-4	&	3.113045E-2	&	0(0){$\rightarrow$}1(1v7,4v8)	\\
			1.0715	&	1.396062E-3	&	3.736391E-2	&	0(0){$\rightarrow$}1(2v9,1v8,2v6)	\\
			1.0722	&	1.317948E-3	&	-3.630355E-2	&	0(0){$\rightarrow$}1(3v6,1v5)	\\
			1.0731	&	8.206839E-3	&	-9.059161E-2	&	0(0){$\rightarrow$}1(1v8,2v6,1v4)	\\
			1.0735	&	2.722161E-3	&	-5.217433E-2	&	0(0){$\rightarrow$}1(1v6,1v2)	\\
			1.0761	&	2.198445E-3	&	-4.688758E-2	&	0(0){$\rightarrow$}1(1v7,3v8,1v6)	\\
			1.0764	&	5.538940E-4	&	-2.353495E-2	&	0(0){$\rightarrow$}1(1v7,1v6,2v4)	\\
			1.0776	&	1.799413E-3	&	4.241949E-2	&	0(0){$\rightarrow$}1(3v8,1v3)	\\
			1.0801	&	7.261782E-4	&	-2.694769E-2	&	0(0){$\rightarrow$}1(2v9,3v6)	\\
			1.0817	&	4.268886E-3	&	6.533671E-2	&	0(0){$\rightarrow$}1(3v6,1v4)	\\
			1.0846	&	2.598904E-3	&	5.097944E-2	&	0(0){$\rightarrow$}1(1v7,2v8,2v6)	\\
			1.0861	&	3.606802E-3	&	-6.005666E-2	&	0(0){$\rightarrow$}1(2v8,1v6,1v3)	\\
			1.0932	&	1.909612E-3	&	-4.369911E-2	&	0(0){$\rightarrow$}1(1v7,1v8,3v6)	\\
			1.0946	&	3.493776E-3	&	5.910818E-2	&	0(0){$\rightarrow$}1(1v8,2v6,1v3)	\\
			1.1010	&	9.558551E-4	&	-3.091691E-2	&	0(0){$\rightarrow$}1(2v8,1v5,1v4)	\\
			1.1017	&	8.300868E-4	&	2.881123E-2	&	0(0){$\rightarrow$}1(1v7,4v6)	\\
			1.1032	&	1.817330E-3	&	-4.263015E-2	&	0(0){$\rightarrow$}1(3v6,1v3)	\\
			1.1089	&	5.266681E-4	&	-2.294925E-2	&	0(0){$\rightarrow$}1(2v9,2v8,1v4)	\\
			1.1096	&	1.569935E-3	&	3.962241E-2	&	0(0){$\rightarrow$}1(1v8,1v6,1v5,1v4)	\\
			1.1102	&	2.339679E-3	&	4.837023E-2	&	0(0){$\rightarrow$}1(5v8)	\\
			1.1105	&	1.737953E-3	&	4.168876E-2	&	0(0){$\rightarrow$}1(2v8,2v4)	\\
			1.1175	&	8.650211E-4	&	2.941124E-2	&	0(0){$\rightarrow$}1(2v9,1v8,1v6,1v4)	\\
			1.1181	&	1.076560E-3	&	-3.281097E-2	&	0(0){$\rightarrow$}1(2v6,1v5,1v4)	\\
			1.1188	&	5.806910E-3	&	-7.620308E-2	&	0(0){$\rightarrow$}1(4v8,1v6)	\\
			1.1191	&	2.854484E-3	&	-5.342738E-2	&	0(0){$\rightarrow$}1(1v8,1v6,2v4)	\\
			1.1194	&	9.947616E-4	&	-3.153984E-2	&	0(0){$\rightarrow$}1(1v4,1v2)	\\
			1.1220	&	8.033797E-4	&	-2.834395E-2	&	0(0){$\rightarrow$}1(1v7,3v8,1v4)	\\
			1.1260	&	5.931754E-4	&	-2.435519E-2	&	0(0){$\rightarrow$}1(2v9,2v6,1v4)	\\
			1.1273	&	7.769142E-3	&	8.814274E-2	&	0(0){$\rightarrow$}1(3v8,2v6)	\\
			1.1276	&	1.957420E-3	&	4.424274E-2	&	0(0){$\rightarrow$}1(2v6,2v4)	\\
			1.1306	&	1.610321E-3	&	4.012880E-2	&	0(0){$\rightarrow$}1(1v7,2v8,1v6,1v4)	\\
			1.1311	&	6.683453E-4	&	-2.585238E-2	&	0(0){$\rightarrow$}1(1v8,1v6,1v5,1v3)	\\
			1.1320	&	1.318037E-3	&	-3.630478E-2	&	0(0){$\rightarrow$}1(2v8,1v4,1v3)	\\
			1.1358	&	6.966744E-3	&	-8.346702E-2	&	0(0){$\rightarrow$}1(2v8,3v6)	\\
			1.1391	&	1.559858E-3	&	-3.949504E-2	&	0(0){$\rightarrow$}1(1v7,1v8,2v6,1v4)	\\
			1.1394	&	5.173959E-4	&	-2.274634E-2	&	0(0){$\rightarrow$}1(1v7,1v6,1v2)	\\
			1.1406	&	2.164797E-3	&	4.652738E-2	&	0(0){$\rightarrow$}1(1v8,1v6,1v4,1v3)	\\
			1.1444	&	4.277839E-3	&	6.540519E-2	&	0(0){$\rightarrow$}1(1v8,4v6)	\\
			1.1476	&	8.113790E-4	&	2.848472E-2	&	0(0){$\rightarrow$}1(1v7,3v6,1v4)	\\
			1.1491	&	1.484478E-3	&	-3.852892E-2	&	0(0){$\rightarrow$}1(2v6,1v4,1v3)	\\
			1.1521	&	6.855380E-4	&	-2.618278E-2	&	0(0){$\rightarrow$}1(1v7,2v8,1v6,1v3)	\\
			1.1529	&	1.633730E-3	&	-4.041943E-2	&	0(0){$\rightarrow$}1(5v6)	\\
			1.1552	&	6.551398E-4	&	2.559570E-2	&	0(0){$\rightarrow$}1(4v8,1v5)	\\
			1.1606	&	6.640553E-4	&	2.576927E-2	&	0(0){$\rightarrow$}1(1v7,1v8,2v6,1v3)	\\
			1.1621	&	5.868952E-4	&	-2.422592E-2	&	0(0){$\rightarrow$}1(1v8,1v6,2v3)	\\
			1.1637	&	1.486206E-3	&	-3.855134E-2	&	0(0){$\rightarrow$}1(3v8,1v6,1v5)	\\
			1.1647	&	2.122024E-3	&	-4.606543E-2	&	0(0){$\rightarrow$}1(4v8,1v4)	\\
			1.1716	&	8.188868E-4	&	-2.861620E-2	&	0(0){$\rightarrow$}1(2v9,3v8,1v6)	\\
			1.1723	&	1.756926E-3	&	4.191570E-2	&	0(0){$\rightarrow$}1(2v8,2v6,1v5)	\\
			1.1732	&	4.813880E-3	&	6.938213E-2	&	0(0){$\rightarrow$}1(3v8,1v6,1v4)	\\
			1.1736	&	5.003111E-4	&	2.236763E-2	&	0(0){$\rightarrow$}1(1v6,3v4)	\\
			1.1736	&	1.623433E-3	&	4.029184E-2	&	0(0){$\rightarrow$}1(2v8,1v2)	\\
			1.1802	&	9.680514E-4	&	3.111352E-2	&	0(0){$\rightarrow$}1(2v9,2v8,2v6)	\\
			1.1808	&	1.290947E-3	&	-3.592975E-2	&	0(0){$\rightarrow$}1(1v8,3v6,1v5)	\\
			1.1818	&	5.690753E-3	&	-7.543708E-2	&	0(0){$\rightarrow$}1(2v8,2v6,1v4)	\\
			1.1821	&	2.666392E-3	&	-5.163712E-2	&	0(0){$\rightarrow$}1(1v8,1v6,1v2)	\\
			1.1847	&	1.103708E-3	&	-3.322210E-2	&	0(0){$\rightarrow$}1(1v7,4v8,1v6)	\\
			1.1850	&	5.425463E-4	&	-2.329262E-2	&	0(0){$\rightarrow$}1(1v7,1v8,1v6,2v4)	\\
			1.1862	&	9.033779E-4	&	3.005625E-2	&	0(0){$\rightarrow$}1(4v8,1v3)	\\
			1.1887	&	7.113009E-4	&	-2.667023E-2	&	0(0){$\rightarrow$}1(2v9,1v8,3v6)	\\
			1.1893	&	5.611601E-4	&	2.368882E-2	&	0(0){$\rightarrow$}1(4v6,1v5)	\\
			1.1903	&	4.181429E-3	&	6.466397E-2	&	0(0){$\rightarrow$}1(1v8,3v6,1v4)	\\
			1.1907	&	1.828439E-3	&	4.276025E-2	&	0(0){$\rightarrow$}1(2v6,1v2)	\\
			1.1933	&	1.476666E-3	&	3.842741E-2	&	0(0){$\rightarrow$}1(1v7,3v8,2v6)	\\
			1.1947	&	2.049342E-3	&	-4.526966E-2	&	0(0){$\rightarrow$}1(3v8,1v6,1v3)	\\
			1.1951	&	5.163277E-4	&	-2.272285E-2	&	0(0){$\rightarrow$}1(1v6,2v4,1v3)	\\
			1.1989	&	1.817620E-3	&	-4.263355E-2	&	0(0){$\rightarrow$}1(4v6,1v4)	\\
			1.2018	&	1.324156E-3	&	-3.638895E-2	&	0(0){$\rightarrow$}1(1v7,2v8,3v6)	\\
			1.2033	&	2.422640E-3	&	4.922032E-2	&	0(0){$\rightarrow$}1(2v8,2v6,1v3)	\\
			1.2097	&	5.431054E-4	&	-2.330462E-2	&	0(0){$\rightarrow$}1(3v8,1v5,1v4)	\\
			1.2103	&	8.130807E-4	&	2.851457E-2	&	0(0){$\rightarrow$}1(1v7,1v8,4v6)	\\
			1.2118	&	1.780098E-3	&	-4.219121E-2	&	0(0){$\rightarrow$}1(1v8,3v6,1v3)	\\
			1.2182	&	1.088618E-3	&	3.299421E-2	&	0(0){$\rightarrow$}1(2v8,1v6,1v5,1v4)	\\
			1.2189	&	1.002857E-3	&	3.166792E-2	&	0(0){$\rightarrow$}1(6v8)	\\
			1.2192	&	9.874839E-4	&	3.142426E-2	&	0(0){$\rightarrow$}1(3v8,2v4)	\\
			1.2204	&	7.737884E-4	&	2.781705E-2	&	0(0){$\rightarrow$}1(4v6,1v3)	\\
			1.2261	&	5.998194E-4	&	2.449121E-2	&	0(0){$\rightarrow$}1(2v9,2v8,1v6,1v4)	\\
			1.2267	&	1.054504E-3	&	-3.247313E-2	&	0(0){$\rightarrow$}1(1v8,2v6,1v5,1v4)	\\
			1.2274	&	2.664648E-3	&	-5.162023E-2	&	0(0){$\rightarrow$}1(5v8,1v6)	\\
			1.2277	&	1.979345E-3	&	-4.448983E-2	&	0(0){$\rightarrow$}1(2v8,1v6,2v4)	\\
			1.2281	&	9.743818E-4	&	-3.121509E-2	&	0(0){$\rightarrow$}1(1v8,1v4,1v2)	\\
			1.2347	&	5.810230E-4	&	-2.410442E-2	&	0(0){$\rightarrow$}1(2v9,1v8,2v6,1v4)	\\
			1.2353	&	5.485131E-4	&	2.342036E-2	&	0(0){$\rightarrow$}1(3v6,1v5,1v4)	\\
			1.2359	&	3.900423E-3	&	6.245336E-2	&	0(0){$\rightarrow$}1(4v8,2v6)	\\
			1.2363	&	1.917319E-3	&	4.378720E-2	&	0(0){$\rightarrow$}1(1v8,2v6,2v4)	\\
			1.2366	&	1.132929E-3	&	3.365901E-2	&	0(0){$\rightarrow$}1(1v6,1v4,1v2)	\\
			1.2392	&	9.149648E-4	&	3.024839E-2	&	0(0){$\rightarrow$}1(1v7,3v8,1v6,1v4)	\\
			1.2407	&	7.488927E-4	&	-2.736590E-2	&	0(0){$\rightarrow$}1(3v8,1v4,1v3)	\\
			1.2445	&	3.958420E-3	&	-6.291598E-2	&	0(0){$\rightarrow$}1(3v8,3v6)	\\
			1.2448	&	9.973163E-4	&	-3.158032E-2	&	0(0){$\rightarrow$}1(3v6,2v4)	\\
			1.2477	&	1.081631E-3	&	-3.288815E-2	&	0(0){$\rightarrow$}1(1v7,2v8,2v6,1v4)	\\
			1.2481	&	5.067960E-4	&	-2.251213E-2	&	0(0){$\rightarrow$}1(1v7,1v8,1v6,1v2)	\\
			1.2492	&	1.501105E-3	&	3.874410E-2	&	0(0){$\rightarrow$}1(2v8,1v6,1v4,1v3)	\\
			1.2530	&	2.966322E-3	&	5.446395E-2	&	0(0){$\rightarrow$}1(2v8,4v6)	\\
			1.2563	&	7.947562E-4	&	2.819142E-2	&	0(0){$\rightarrow$}1(1v7,1v8,3v6,1v4)	\\
			1.2578	&	1.454065E-3	&	-3.813221E-2	&	0(0){$\rightarrow$}1(1v8,2v6,1v4,1v3)	\\
			1.2616	&	1.600260E-3	&	-4.000325E-2	&	0(0){$\rightarrow$}1(1v8,5v6)	\\
			1.2663	&	7.563495E-4	&	2.750181E-2	&	0(0){$\rightarrow$}1(3v6,1v4,1v3)	\\
			1.2701	&	5.533435E-4	&	2.352325E-2	&	0(0){$\rightarrow$}1(6v6)	\\
			1.2724	&	7.461351E-4	&	-2.731547E-2	&	0(0){$\rightarrow$}1(4v8,1v6,1v5)	\\
			1.2733	&	9.737446E-4	&	-3.120488E-2	&	0(0){$\rightarrow$}1(5v8,1v4)	\\
			1.2809	&	9.982642E-4	&	3.159532E-2	&	0(0){$\rightarrow$}1(3v8,2v6,1v5)	\\
			1.2819	&	2.416762E-3	&	4.916057E-2	&	0(0){$\rightarrow$}1(4v8,1v6,1v4)	\\
			1.2822	&	9.224149E-4	&	3.037128E-2	&	0(0){$\rightarrow$}1(3v8,1v2)	\\
			1.2888	&	5.500352E-4	&	2.345283E-2	&	0(0){$\rightarrow$}1(2v9,3v8,2v6)	\\
			1.2894	&	8.951633E-4	&	-2.991928E-2	&	0(0){$\rightarrow$}1(2v8,3v6,1v5)	\\
			1.2904	&	3.233418E-3	&	-5.686315E-2	&	0(0){$\rightarrow$}1(3v8,2v6,1v4)	\\
			1.2908	&	1.848919E-3	&	-4.299905E-2	&	0(0){$\rightarrow$}1(2v8,1v6,1v2)	\\
			1.2934	&	5.064645E-4	&	-2.250477E-2	&	0(0){$\rightarrow$}1(1v7,5v8,1v6)	\\
			1.2980	&	5.496635E-4	&	2.344490E-2	&	0(0){$\rightarrow$}1(1v8,4v6,1v5)	\\
			1.2990	&	2.899470E-3	&	5.384672E-2	&	0(0){$\rightarrow$}1(2v8,3v6,1v4)	\\
			1.2993	&	1.790979E-3	&	4.231996E-2	&	0(0){$\rightarrow$}1(1v8,2v6,1v2)	\\
			1.3019	&	7.413458E-4	&	2.722767E-2	&	0(0){$\rightarrow$}1(1v7,4v8,2v6)	\\
			1.3034	&	1.028852E-3	&	-3.207573E-2	&	0(0){$\rightarrow$}1(4v8,1v6,1v3)	\\
			1.3037	&	5.057496E-4	&	-2.248888E-2	&	0(0){$\rightarrow$}1(1v8,1v6,2v4,1v3)	\\
			1.3075	&	1.780382E-3	&	-4.219457E-2	&	0(0){$\rightarrow$}1(1v8,4v6,1v4)	\\
			1.3078	&	9.315995E-4	&	-3.052211E-2	&	0(0){$\rightarrow$}1(3v6,1v2)	\\
			1.3105	&	7.523693E-4	&	-2.742935E-2	&	0(0){$\rightarrow$}1(1v7,3v8,3v6)	\\
			1.3119	&	1.376515E-3	&	3.710142E-2	&	0(0){$\rightarrow$}1(3v8,2v6,1v3)	\\
			1.3161	&	6.799376E-4	&	2.607561E-2	&	0(0){$\rightarrow$}1(5v6,1v4)	\\
			1.3190	&	5.638031E-4	&	2.374454E-2	&	0(0){$\rightarrow$}1(1v7,2v8,4v6)	\\
			1.3205	&	1.234348E-3	&	-3.513329E-2	&	0(0){$\rightarrow$}1(2v8,3v6,1v3)	\\
			1.3268	&	6.185398E-4	&	2.487046E-2	&	0(0){$\rightarrow$}1(3v8,1v6,1v5,1v4)	\\
			1.3290	&	7.579358E-4	&	2.753063E-2	&	0(0){$\rightarrow$}1(1v8,4v6,1v3)	\\
			1.3354	&	7.312101E-4	&	-2.704090E-2	&	0(0){$\rightarrow$}1(2v8,2v6,1v5,1v4)	\\
			1.3361	&	1.142149E-3	&	-3.379569E-2	&	0(0){$\rightarrow$}1(6v8,1v6)	\\
			1.3364	&	1.124640E-3	&	-3.353565E-2	&	0(0){$\rightarrow$}1(3v8,1v6,2v4)	\\
			1.3367	&	6.756519E-4	&	-2.599331E-2	&	0(0){$\rightarrow$}1(2v8,1v4,1v2)	\\
			1.3439	&	5.372757E-4	&	2.317921E-2	&	0(0){$\rightarrow$}1(1v8,3v6,1v5,1v4)	\\
			1.3446	&	1.789808E-3	&	4.230612E-2	&	0(0){$\rightarrow$}1(5v8,2v6)	\\
			1.3449	&	1.329499E-3	&	3.646230E-2	&	0(0){$\rightarrow$}1(2v8,2v6,2v4)	\\
			1.3452	&	1.109718E-3	&	3.331243E-2	&	0(0){$\rightarrow$}1(1v8,1v6,1v4,1v2)	\\
			1.3531	&	1.987287E-3	&	-4.457899E-2	&	0(0){$\rightarrow$}1(4v8,3v6)	\\
			1.3535	&	9.768842E-4	&	-3.125515E-2	&	0(0){$\rightarrow$}1(1v8,3v6,2v4)	\\
			1.3538	&	7.609728E-4	&	-2.758574E-2	&	0(0){$\rightarrow$}1(2v6,1v4,1v2)	\\
			1.3564	&	6.145695E-4	&	-2.479051E-2	&	0(0){$\rightarrow$}1(1v7,3v8,2v6,1v4)	\\
			1.3579	&	8.529099E-4	&	2.920462E-2	&	0(0){$\rightarrow$}1(3v8,1v6,1v4,1v3)	\\
			1.3617	&	1.685429E-3	&	4.105397E-2	&	0(0){$\rightarrow$}1(3v8,4v6)	\\
			1.3649	&	5.510966E-4	&	2.347545E-2	&	0(0){$\rightarrow$}1(1v7,2v8,3v6,1v4)	\\
			1.3664	&	1.008272E-3	&	-3.175330E-2	&	0(0){$\rightarrow$}1(2v8,2v6,1v4,1v3)	\\
			1.3702	&	1.109646E-3	&	-3.331134E-2	&	0(0){$\rightarrow$}1(2v8,5v6)	\\
			1.3750	&	7.408540E-4	&	2.721863E-2	&	0(0){$\rightarrow$}1(1v8,3v6,1v4,1v3)	\\
			1.3788	&	5.420071E-4	&	2.328105E-2	&	0(0){$\rightarrow$}1(1v8,6v6)	\\
			1.3896	&	5.011689E-4	&	2.238680E-2	&	0(0){$\rightarrow$}1(4v8,2v6,1v5)	\\
			1.3905	&	1.108992E-3	&	3.330154E-2	&	0(0){$\rightarrow$}1(5v8,1v6,1v4)	\\
			1.3981	&	5.086210E-4	&	-2.255263E-2	&	0(0){$\rightarrow$}1(3v8,3v6,1v5)	\\
			1.3991	&	1.623306E-3	&	-4.029027E-2	&	0(0){$\rightarrow$}1(4v8,2v6,1v4)	\\
			1.3994	&	1.050533E-3	&	-3.241193E-2	&	0(0){$\rightarrow$}1(3v8,1v6,1v2)	\\
			1.4076	&	1.647444E-3	&	4.058872E-2	&	0(0){$\rightarrow$}1(3v8,3v6,1v4)	\\
			1.4079	&	1.241894E-3	&	3.524051E-2	&	0(0){$\rightarrow$}1(2v8,2v6,1v2)	\\
			1.4162	&	1.234545E-3	&	-3.513610E-2	&	0(0){$\rightarrow$}1(2v8,4v6,1v4)	\\
			1.4165	&	9.125137E-4	&	-3.020784E-2	&	0(0){$\rightarrow$}1(1v8,3v6,1v2)	\\
			1.4206	&	6.910661E-4	&	2.628814E-2	&	0(0){$\rightarrow$}1(4v8,2v6,1v3)	\\
			1.4247	&	6.660076E-4	&	2.580712E-2	&	0(0){$\rightarrow$}1(1v8,5v6,1v4)	\\
			1.4291	&	7.013419E-4	&	-2.648286E-2	&	0(0){$\rightarrow$}1(3v8,3v6,1v3)	\\
			1.4377	&	5.255648E-4	&	2.292520E-2	&	0(0){$\rightarrow$}1(2v8,4v6,1v3)	\\
			1.4450	&	5.646147E-4	&	-2.376162E-2	&	0(0){$\rightarrow$}1(4v8,1v6,2v4)	\\
			1.4532	&	7.671659E-4	&	2.769776E-2	&	0(0){$\rightarrow$}1(6v8,2v6)	\\
			1.4536	&	7.554055E-4	&	2.748464E-2	&	0(0){$\rightarrow$}1(3v8,2v6,2v4)	\\
			1.4539	&	7.694963E-4	&	2.773980E-2	&	0(0){$\rightarrow$}1(2v8,1v6,1v4,1v2)	\\
			1.4618	&	9.119169E-4	&	-3.019796E-2	&	0(0){$\rightarrow$}1(5v8,3v6)	\\
			1.4621	&	6.773871E-4	&	-2.602666E-2	&	0(0){$\rightarrow$}1(2v8,3v6,2v4)	\\
			1.4624	&	7.453827E-4	&	-2.730170E-2	&	0(0){$\rightarrow$}1(1v8,2v6,1v4,1v2)	\\
			1.4703	&	8.461531E-4	&	2.908871E-2	&	0(0){$\rightarrow$}1(4v8,4v6)	\\
			1.4751	&	5.728880E-4	&	-2.393508E-2	&	0(0){$\rightarrow$}1(3v8,2v6,1v4,1v3)	\\
			1.4789	&	6.304873E-4	&	-2.510951E-2	&	0(0){$\rightarrow$}1(3v8,5v6)	\\
			1.4836	&	5.137200E-4	&	2.266539E-2	&	0(0){$\rightarrow$}1(2v8,3v6,1v4,1v3)	\\
			1.5077	&	7.448952E-4	&	-2.729277E-2	&	0(0){$\rightarrow$}1(5v8,2v6,1v4)	\\
			1.5081	&	5.274101E-4	&	-2.296541E-2	&	0(0){$\rightarrow$}1(4v8,1v6,1v2)	\\
			1.5163	&	8.270832E-4	&	2.875905E-2	&	0(0){$\rightarrow$}1(4v8,3v6,1v4)	\\
			1.5166	&	7.056291E-4	&	2.656368E-2	&	0(0){$\rightarrow$}1(3v8,2v6,1v2)	\\
			1.5248	&	7.014539E-4	&	-2.648497E-2	&	0(0){$\rightarrow$}1(3v8,4v6,1v4)	\\
			1.5251	&	6.327516E-4	&	-2.515455E-2	&	0(0){$\rightarrow$}1(2v8,3v6,1v2)	\\
			1.5711	&	5.168603E-4	&	-2.273456E-2	&	0(0){$\rightarrow$}1(2v8,2v6,1v4,1v2)	\\		
	\end{longtable}
\end{center}

\end{document}